%% file: paper.tex
\documentclass[sigconf]{acmart}

\copyrightyear{2026}
\acmYear{2026}
\setcopyright{cc}
\setcctype{by}
\acmConference[CIKM '26]{Proceedings of the 35th ACM International Conference on Information and Knowledge Management}{November 07--11, 2026}{Rome, Italy}
\acmBooktitle{Proceedings of the 35th ACM International Conference on Information and Knowledge Management (CIKM '26), November 07--11, 2026, Rome, Italy}
\acmDOI{10.1145/3799682.3841064}
\acmISBN{979-8-4007-2539-5/2026/11}

\usepackage{amsmath,amsthm}
\usepackage{algorithm,algorithmicx}
\usepackage{algpseudocode}
\usepackage{enumitem,graphicx}
\usepackage{subcaption}
\usepackage{xspace} 
\usepackage{cancel} 
\usepackage{caption}
\input{macros}

\begin{document}

\title{Approximate Pareto Frontiers for Submodular Utility and Cost Tradeoffs}

\author{Karan Vombatkere}
\affiliation{%
  \institution{Boston University}
  \city{Boston}
  \state{MA}
  \country{USA}}
\email{kvombat@bu.edu}

\author{Evimaria Terzi}
\affiliation{%
  \institution{Boston University}
  \city{Boston}
  \state{MA}
  \country{USA}}
\email{evimaria@bu.edu}

\renewcommand{\shortauthors}{Karan Vombatkere and Evimaria Terzi}

\begin{abstract}
In many data-mining applications, including recommender systems, influence maximization, and team formation, the goal is to pick
a subset of elements (e.g., items, nodes in a network, experts to perform a task) to maximize
a monotone submodular utility function while simultaneously minimizing a cost function.
Classical formulations model this tradeoff via cardinality or knapsack constraints, or by combining utility and cost into a single weighted objective. However, such approaches require committing to a specific tradeoff in advance and return only a single solution, offering limited insight into the space of viable utility–cost tradeoffs.

In this paper, we depart from the single-solution paradigm and examine the problem of computing 
representative sets of high-quality solutions that expose different tradeoffs between submodular utility and cost.
For this, we introduce $(\alpha_1,\alpha_2)$-approximate Pareto
frontiers that provably approximate the achievable tradeoffs between submodular utility and cost.
Specifically, we formalize the \textsc{Pareto--}$\langle f,c \rangle$ problem and develop efficient algorithms for multiple instantiations arising from different combinations of submodular utility $f$ and cost functions $c$.
We also provide an adaptive search algorithm that computes only a small subset of points that collectively summarize the entire Pareto frontier.

Our results offer a principled and practical framework for understanding and exploiting utility--cost tradeoffs in submodular optimization. 
Experiments on datasets from diverse application
domains demonstrate that our algorithms efficiently compute approximate Pareto frontiers in practice.
\end{abstract}



\maketitle

\input{1introduction}\label{sec:intro}
\input{2related}\label{sec:related}
\input{3preliminaries}\label{sec:preliminaries}
\input{4problem}\label{sec:problem}
\input{5algorithms}\label{sec:algorithms}
\input{6experiments}\label{sec:experiments}
\input{7conclusion}\label{conclusion}


\clearpage
\section*{GenAI Usage Disclosure}
Generative AI tools (specifically, Claude and Codex) were used to assist with manuscript editing and to accelerate routine coding tasks during experimentation. All research design, algorithmic contributions, analysis, and conclusions are the sole work of the authors.


\bibliographystyle{ACM-Reference-Format}
\bibliography{references}

\end{document}

%% file: macros.tex
\newtheorem{definition}{Definition}
\newtheorem{problem}{Problem}

\newtheorem{proposition}{Proposition}

\newcommand{\paretosubmodcost}{\textsc{Pareto--}\ensuremath{\langle f,c \rangle}\xspace}
\newcommand{\paretosubmodcardinality}{\textsc{Pareto--}\ensuremath{\langle f,\cardinality \rangle}\xspace}
\newcommand{\paretosubmodknapsack}{\textsc{Pareto--}\ensuremath{\langle f,\linear \rangle}\xspace}
\newcommand{\paretosubmoddiam}{\textsc{Pareto--}\ensuremath{\langle f,\diameter \rangle}\xspace}

\newcommand{\bigO}{\ensuremath{\mathcal{O}}\xspace}

\newcommand{\calP}{\ensuremath{\mathcal{P}}\xspace}

\newcommand{\groundset}{\mathcal{V}}

\newcommand{\cardinality}{\ensuremath{\texttt{Cardinality}}}
\newcommand{\linear}{\ensuremath{\texttt{Linear}}}
\newcommand{\diameter}{\ensuremath{\texttt{Diameter}}}

\newcommand{\freelancer}{{\textit{Freelancer}}\xspace}
\newcommand{\bibsonomy}{\textit{Bibsonomy}\xspace}
\newcommand{\imdb}{{\textit{IMDB}}\xspace}
\newcommand{\imdbone}{{{\imdb}\textit{-1}}\xspace}
\newcommand{\imdbtwo}{{{\imdb}\textit{-2}}\xspace}
\newcommand{\yelpvegas}{{\textit{Yelp-LV}}\xspace}
\newcommand{\yelpphoenix}{{\textit{Yelp-PHX}}\xspace}
\newcommand{\NetHEPT}{{\textit{NetHEPT}}\xspace}
\newcommand{\NetPHY}{{\textit{NetPHY}}\xspace}

\newcommand{\greedy}{\normalfont{\texttt{Greedy}}\xspace}
\newcommand{\cgreedy}{\normalfont{\texttt{C-Greedy}}\xspace}

\newcommand{\cgreedydiameter}{\normalfont{\texttt{C-Greedy-Diam}}\xspace}

\newcommand{\fgreedy}{\normalfont{\texttt{F-Greedy}}\xspace}

\newcommand{\fcgreedy}{\normalfont{\texttt{FC-Greedy}}\xspace}

\newcommand{\baselinetopk}{\normalfont{\texttt{Top\-K}}\xspace}
\newcommand{\baselinerandom}{\normalfont{\texttt{Random}}\xspace}

\newcommand{\prunegraph}{\normalfont{\texttt{Prune\-Graph}}\xspace}
\newcommand{\topkdegree}{\normalfont{\texttt{Top\-Degree}}\xspace}
\newcommand{\distancegreedy}{\normalfont{\texttt{DistanceGreedy}}\xspace}

\newcommand{\paretogreedy}{\normalfont{\texttt{Pareto\-Greedy}}\xspace}
\newcommand{\paretosummary}{\normalfont{\texttt{Pareto\-Summary}}\xspace}

\newcommand{\spara}[1]{\vspace{0.3em}\noindent{\bf{#1}}}

\newcommand{\etal}{{et al.}}
\newcommand{\ie}{{i.e.}}

%% file: 1introduction.tex
\section{Introduction}
Submodular optimization appears in many data-mining applications, including
\emph{recommender systems}~\cite{el2009turning,tschiatschek2017selecting,kazemi2021regularized},
\emph{influence maximization}~\cite{kempe2003maximizing,chen2009efficient}, and
\emph{team formation}~\cite{nikolakaki21efficient,vombatkere2023balancing,vombatkere2025forming,vombatkere2025qubo}.
In these settings, the goal is to select a set of entities (e.g., items, nodes in a social network, or experts in team formation)
that maximizes a monotone submodular utility function while minimizing an associated cost.
The specific instantiations of both the utility and cost functions are application-dependent; for example, in influence maximization,
the objective is to maximize the spread of influence while minimizing the cost of seeding nodes, whereas in team formation the goal
is to cover required skills while minimizing coordination costs.
Traditionally, such problems have been studied under cardinality or matroid constraints,
for which strong approximation guarantees are known~\cite{nemhauser1978analysis,krause2014submodular,
sviridenko2004note,khuller1999budgeted,calinescu2011maximizing,kulik2013approximations}.

More recent work has considered multi-objective formulations that combine submodular utility and cost into a single weighted objective~\cite{harshaw2019submodular,kazemi2021regularized,
nikolakaki21efficient,vombatkere2023balancing,vombatkere2025forming,vombatkere2025qubo}.
While this approach is flexible, it requires selecting weights \emph{a priori} and typically produces only a single solution.
Moreover, the resulting objective may take both positive and negative values, for which standard approximation guarantees no longer apply,
and small changes in the weights can lead to qualitatively different solutions.

A common characteristic of the above approaches is their focus on computing a \emph{single} solution.
In many real-world decision-making scenarios, this single-solution paradigm is limiting.
Practitioners are often interested in understanding the broader tradeoff between utility and cost:
budgets may be negotiable, different stakeholders may prefer different operating points, and small increases in cost may yield
disproportionately large gains in utility.
Returning a single solution obscures these tradeoffs and makes it difficult to reason about alternative solutions. 

In this paper, we depart from the single-solution paradigm and study utility--cost tradeoffs in submodular maximization through the lens of Pareto optimization. This conceptual shift to Pareto frontier computation provides a foundation for future work that is not captured by existing formulations, and consequently enables principled reasoning over the entire utility–cost tradeoff space for a wide variety of data-mining applications. 

We formalize the \paretosubmodcost\ problem, with the goal of computing a \emph{polynomial-size}, \emph{representative} set of solutions that approximately captures the Pareto frontier between submodular utility $f$ and cost $c$. 
We introduce $(\alpha_1,\alpha_2)$-approximate Pareto frontiers tailored to submodular maximization and develop efficient algorithms for finding them.
Our algorithms work for arbitrary monotone submodular utility functions and a broad class of cost functions, including cardinality, knapsack, and graph-based costs, modeling problems that appear in recommender systems, influence maximization, and team formation.

%% file: 2related.tex
\section{Related Work}

\spara{Submodular Maximization with Constraints.}
Maximizing a monotone submodular function under constraints is a central problem
in combinatorial optimization, with well-established guarantees.
Under a cardinality constraint, the greedy algorithm achieves the optimal
$(1-1/e)$ approximation~\cite{nemhauser1978analysis,krause2014submodular}.
This guarantee extends to knapsack and matroid constraints via techniques such as partial enumeration and continuous relaxations
\cite{sviridenko2004note,khuller1999budgeted,calinescu2011maximizing,
kulik2013approximations}.
Subsequent work has focused on improving efficiency and robustness,
leading to fast greedy variants, lazy evaluations, and guess-free algorithms that retain strong guarantees~\cite{badanidiyuru2014fast,feldman2021guess,
feldman2023practical}.
These methods are widely used in data-mining applications where costs correspond
to budgets or capacities, including influence maximization
\cite{kempe2003maximizing,chen2009efficient}, sensor placement
\cite{krause2006near}, and recommendation systems
\cite{kazemi2021regularized}.
A parallel line of work considers submodular maximization under richer structural constraints, particularly tree or graph-based costs.
For example, work in team formation considers expert
communication costs~\cite{lappas2009finding,anagnostopoulos10power,
anagnostopoulos2012online,gajewar2012multi}.
In all cases, the goal is to compute a single solution that optimizes the objective under constraints.

\spara{Combined Objective Treatment.}
An alternative to constrained submodular maximization is to combine utility and
cost into a single objective, typically via weighted sums or regularization~\cite{harshaw2019submodular,mitra2021submodular,zhu2024regularized,guo2026efficient,cui2023constrained}.

This paradigm has been widely adopted in recommendation systems that incorporate cost or diversity
penalties directly into the objective~\cite{kazemi2021regularized}, and team
formation, where hiring costs, workload balance, or social compatibility are
modeled as part of a unified objective
\cite{nikolakaki20finding,nikolakaki21efficient,vombatkere2023balancing,
vombatkere2025forming,vombatkere2025qubo}.
Such formulations admit efficient algorithms, but they require
fixing tradeoff weights \emph{a priori} and they return a \emph{single solution},
obscuring the underlying utility--cost tradeoffs and limit the ability to reason about alternative solutions.

Our work departs from the single-solution paradigm and formalizes the
submodular utility--cost tradeoff by approximating the Pareto frontier across 
cost models.

\spara{Approximate Pareto Frontiers.}
Approximate Pareto frontiers originate in the work of Papadimitriou and
Yannakakis~\cite{papadimitriou2000approximability}, who showed that any
multi-objective optimization problem admits a polynomial-size
$\varepsilon$-approximate Pareto set.
Subsequent work developed methods for constructing succinct tradeoff
curves in combinatorial optimization~\cite{vassilvitskii2005efficiently,
diakonikolas2010small}.
These results are largely existential and typically assume access to oracles or near-exact solvers.

Algorithmic developments in this area have primarily focused on linear or
problem-specific objectives and rely on structural assumptions that do not
extend to general submodular functions.
Representative applications include bicriteria shortest paths
\cite{shekelyan2014linear,galbrun2014safe}, network design
\cite{krause2006near}, and hydropower dam placement
\cite{wu2018efficiently,bai2023efficiently}.
In team formation, skyline and two-phase methods balance coordination quality and
hiring costs~\cite{zihayat2014two,kargar2012efficient,kargar2013finding}, but under complete task-coverage constraints and often without
approximation guarantees.

Additional work explores Pareto tradeoffs in specific domains such as fairness
or quality in clustering and recommendation systems.
For example, Pareto formulations have been used to balance fairness and
clustering quality~\cite{hakim2024fairness} and to preserve diversity and quality
in group recommendation~\cite{xiao2017fairness}.
Learning-based methods aim to approximate Pareto sets empirically, but lack
theoretical guarantees and robustness across cost models
\cite{lin2022pareto,chen2023neural}.

Within submodular optimization, Pareto formulations have received limited
attention. Fazzone et al.~\cite{fazzone2024fair} study the Submodular
Maximization with Fair Representation (SMFR) problem, which jointly maximizes a
primary submodular utility function and multiple representativeness objectives
under knapsack or matroid constraints.
Although minimizing cost is mathematically equivalent to maximizing the
negative cost function, exactly because of this,
the approximation guarantees of their framework cannot be directly applied to our setting.
Soma and Yoshida~\cite{soma2017regret} study representative selection for
multi-objective submodular maximization via regret-ratio minimization, producing
a finite set of solutions corresponding to different non-negative linear
combinations of the objectives.
While this framework captures tradeoffs among multiple submodular functions, it doesn't provide approximation guarantees for individual objectives, and does not yield explicit guarantees on the utility--cost Pareto frontier.

Our work complements this literature by establishing provable approximate Pareto guarantees for monotone submodular objectives under multiple cost models, and by demonstrating their practical effectiveness.

%% file: 3preliminaries.tex
\section{Preliminaries}
Let \(\groundset = \{1,\dots,n\}\) denote a finite ground set of items of cardinality $n$.
A \emph{solution} is any subset \(S \subseteq \groundset\).
We evaluate each solution $S$ using two criteria: the utility function \(f(S)\),
which we seek to maximize, and a cost function \(c(S)\), which we seek to minimize.
We consider the following utility and cost functions.

\spara{Submodular Utilities.}
Throughout the paper, we assume that our utility function  \(f : 2^\groundset \to \mathbb{R}_{\ge 0}\) is a non-negative,
monotone, and submodular function.
Monotonicity implies that 
\begin{align*}
f(S) \le f(T) \quad \forall S \subseteq T \subseteq \groundset.
\end{align*}
Submodularity implies that for all \(S \subseteq T \subseteq \groundset\) and \(u \in \groundset \setminus T\),
\(f\) satisfies the diminishing-returns property:
\begin{align*}
f(T \cup\{u\}) - f(T) \le f(S \cup\{u\}) - f(S).
\end{align*}

\spara{Cost Functions.}
Each solution \(S \subseteq \groundset\) is associated with a non-negative cost
\(c : 2^\groundset \to \mathbb{R}_{\ge 0}\).
The cost function encodes structural or feasibility constraints and may take
different forms. Motivated by work in recommender systems~\cite{kazemi2021regularized}, influence maximization~\cite{kempe2003maximizing} and team formation~\cite{lappas2009finding,nikolakaki21efficient}, we focus here on three cost variants.

\emph{{\cardinality} cost.}
Maximizing a submodular function subject to a cardinality constraint has been well-studied~\cite{krause2014submodular}. To constrain the number of selected items to be at most \(k\), we define the cost function as follows:
\[
c_k(S) = |S|.
\]
This cost balances the \emph{size} of a solution with its utility.

\emph{{\linear} cost.}
Similar to prior work in budgeted submodular maximization and team
formation~\cite{nikolakaki21efficient,khuller1999budgeted,feldman2023practical}, each item \(i \in \groundset\) is assigned a non-negative weight \(w_i\), and the total cost is additive:
\[
c_{\ell}(S) = \sum_{i \in S} w_i.
\]
This is a linear (knapsack-like) cost on the items in \(S\).

\emph{{\diameter} cost.}
We assume an underlying weighted graph \(G = (\groundset,E)\), where edge weights encode pairwise coordination costs~\cite{anagnostopoulos2012online,lappas2009finding,vombatkere2025forming}.
Let \(d(i,j) \ge 0\) denote the shortest-path distance between items
\(i\) and \(j\) in \(G\). We then define the diameter cost of solution \(S\):
\[
c_d(S) = \max_{i,j \in S} d(i,j).
\]
This cost captures the worst-case coordination difficulty within the selected
set and induces a non-linear, non-additive constraint.

\spara{Greedy Primitive.} 
Given a monotone submodular function $f$ and a linear cost function $c$ defined
over a ground set $\groundset$, the \greedy\ procedure constructs a solution
incrementally. It repeatedly adds the feasible element with
maximum cost-scaled marginal gain
\[
\Delta_f(e \mid S)
\;=\;
\frac{f(S \cup \{e\}) - f(S)}{c(\{e\})} .
\]

The notion of feasibility depends on the objective being optimized.
We consider two complementary stopping conditions.
In the \emph{cost-constrained} case, the procedure grows the solution until its
total cost reaches a prescribed budget $B$.
In the \emph{utility-target} case, the procedure continues until the solution
achieves a target utility value $K$.
By instantiating the greedy primitive with either a cost threshold $B$ or a
utility target $K$, we capture both optimization regimes within a single
algorithmic template.

Often, the greedy procedure is initialized with a set of
\emph{seeds}, where each seed is a feasible subset $S_0 \subseteq \groundset$ of
size at most $\tau$; $\tau$ controls the tradeoff between running time and
approximation quality.
Enumerating all such seeds mitigates sensitivity to poor initializations, a
standard technique in submodular maximization~\cite{sviridenko2004note,feldman2023practical}.

\begin{algorithm}
\caption{\greedy $(f,c,K,B,\tau)$}
\label{alg:generic-greedy}
\begin{algorithmic}[1]
\Statex \textbf{Input:} Ground set $\groundset$, monotone submodular $f$, cost $c$, utility target $K$, cost threshold $B$, seed size $\tau$
\State $S^\ast \gets \emptyset$
\ForAll{seed sets $S_0 \subseteq \groundset$ with $|S_0|\le\tau$ and $c(S_0)\le B$}
    \State $S \gets S_0$
    \While{$f(S) \le K$ \text{and} $\exists e \in \groundset \setminus S$ \text{with }$c(S \cup \{e\}) \le B$}
        \State $e^\star \gets 
        \arg\max_{e \in \groundset \setminus S}\Delta_f(e \mid S)$
        \State $S \gets S \cup \{e^\star\}$
    \EndWhile
    \State $S^\ast \gets \arg\max\{f(S^\ast), f(S)\}$
\EndFor
\State \Return $S^\ast$
\end{algorithmic}
\end{algorithm}

\spara{Pareto Frontiers.}
\emph{Pareto points} are a standard concept in multi-objective optimization. In our case, we have two objectives, where one is maximized ($f(\cdot)$) and the other is minimized ($c(\cdot)$).
We describe each problem with the pair $\langle f,c \rangle$ of its objectives. We also describe a solution $S$ by the $2$-dimensional point of the values for these objectives, i.e.~, $(f(S),c(S))$.

\begin{definition}[Pareto Point]
A \emph{Pareto point} is defined as a solution $S\subseteq \groundset$ such that there 
exists  \emph{no}
other solution $S'\subseteq \groundset$ that is better in \emph{both} objectives, {\ie}  there is no other solution $S'$ that satisfies:
\begin{align}
f(S') \geq f(S) \quad \text{and} \quad c(S') \leq c(S).
\end{align}
\end{definition}
In other words, a Pareto point is a solution whose vector of the various objectives (in our case two) is not dominated by the vector of another solution.
The set of all Pareto points 
define the \emph{Pareto Frontier} of the problem and, intuitively, they describe the tradeoffs that different solutions attain.

Note that the Pareto frontier of a multi-objective problem (such as $\langle f,c \rangle$) can have exponential size. 
This resulted in a large body of work on finding \emph{approximate Pareto frontiers}. 
We discuss approximate Pareto frontiers and provide formal definitions related to our problem in the next section.


%% file: 4problem.tex
\section{The {\paretosubmodcost} Problem}
We motivate our definition by considering the Pareto frontier of the bi-objective problem
$\langle f,\cardinality\rangle$,
where the goal is to maximize a submodular function and minimize the {\cardinality} cost function.
Note that since $c$ takes integer values in $\{1,\ldots ,n\}$, the Pareto frontier is polynomial in size. However, in order to find it, we need to solve the following problem for different values of $k$:
\begin{align*}
\max_{S\subseteq \groundset}f(S) \text{ such that } |S|\leq k.
\end{align*}
However, this problem is submodular maximization under cardinality constraints and it is $NP$-hard~\cite{nemhauser1978analysis,vazirani2001approximation}. Therefore, despite the fact that the Pareto frontier is polynomial in size, we cannot find the points in the frontier in polynomial time unless $P=NP$.

 \subsection{Problem Definition}
 For problems like the one above, we need to define \emph{approximate Pareto frontiers}.
 While on a high level, the definition of approximate Pareto frontiers we give is similar to the one proposed by
 Papadimitriou and Yannakakis~\cite{papadimitriou2000approximability}, the motivation behind this definition is different; in their case, the motivation was the size of the frontiers.  In our case, the computational complexity of the underlying computational problem forces us to consider approximate Pareto frontiers regardless of the size of the optimal Pareto frontiers.
We propose the following definition of approximate Pareto frontiers:

\begin{definition}[$(\alpha_1,\alpha_2)$-approximate Pareto frontier]\label{definition:approximate-pareto}
Assume a bi-objective optimization problem $\langle f,c \rangle$
and let 
$\calP^\ast$ be its optimal Pareto frontier for a given input.
For \(\alpha_1 \in (0,1]\) and \(\alpha_2 \ge 1\), we say that a set of solutions $\calP_{\alpha_1,\alpha_2}$ is an $(\alpha_1,\alpha_2)$--approximate Pareto frontier for $\langle f,c \rangle$ such that for every solution $S^\ast\in\calP^\ast$ there exists a  $S\in \calP_{\alpha_1,\alpha_2}$ such that 
\begin{align*}
f(S) \ge \alpha_1 f(S^\ast)
\quad \text{and} \quad
c(S) \le \alpha_2 c(S^\ast).
\end{align*}
\end{definition}
The notion of an $(\alpha_1,\alpha_2)$--approximate Pareto frontier
provides a multiplicative approximation of the Pareto frontier. The closer the values of $\alpha_1$ and $\alpha_2$ are to $1$, the better the approximation of the optimal Pareto frontier. 

The main question is whether there exist values of $\alpha_1$ and $\alpha_2$ such that, for any instance $\langle f,c \rangle$, the corresponding $(\alpha_1,\alpha_2)$-approximate Pareto frontier is both polynomially bounded in size and efficiently computable.
We formalize this in the following definition. 

\begin{problem}[\paretosubmodcost]
Consider a ground set $\groundset$ of \(n\) items, a non-negative monotone submodular function \(f:2^\groundset\to\mathbb{R}_{\ge 0}\) to be maximized, and a non-negative cost function \(c:2^\groundset\to\mathbb{R}_{\ge 0}\) to be minimized.
The goal of \paretosubmodcost\ is to find a polynomial-size set of solutions \(\mathcal{S}' \subseteq 2^\groundset\) that forms an \((\alpha_1,\alpha_2)\)-approximate Pareto frontier for the $\langle f,c\rangle$ problem.
\end{problem}
The \paretosubmodcost problem takes different instantiations depending on the
submodular function $f$ and the cost function $c$ we consider. Throughout, we will assume that $f$ is a monotone submodular function and that $c$ is any of the three cost functions defined in Section~\ref{sec:preliminaries}, {\ie}, {\cardinality}, {\linear} and {\diameter}.





\subsection{Real-World Instantiations}
\label{sec:applications}

We map the \paretosubmodcost\ problem to real-world applications in team formation~\cite{anagnostopoulos10power,anagnostopoulos2012online,lappas2009finding,nikolakaki20finding,vombatkere2023balancing,vombatkere2025forming}, influence maximization~\cite{kempe2003maximizing,chen2009efficient} and  recommender systems~\cite{tschiatschek2017selecting,kazemi2021regularized}. 
We provide formal definitions of \(V, f\), and \(c \), for each case.

\spara{Team Formation.}
The Pareto frontier formulation captures trade-offs between task coverage and team costs in team formation.
Let \(V\) be a set of experts, and let $U$ be a universe of skills.
Each expert \(i\in V\) possesses a subset of skills \(S_i\subseteq U\). 
Given a task $T \subseteq U$, which requires a certain set of skills, the \emph{coverage} of a set of experts $Q \subseteq V$ is a monotone, submodular function defined as:
\begin{align}\label{eqn:cov-objective}
    f(Q)= \big|\; \big(\cup_{i\in Q} S_i \big) \cap T\;\big|.
\end{align}
To encode the linear cost function, we associate a \emph{hiring} cost $w_i$ with each expert. To encode the diameter cost function we use a coordination graph, \(G=(V,E)\) with edge weights \(d(i,j)\) encoding pairwise communication costs between the experts.

\spara{Recommender Systems.}
Following Kazemi et al.~\cite{kazemi2021regularized}, we define the ground set \(V\) to be restaurants and balance recommendation quality against the size or distance between recommended restaurants.
Each item \(i \in V\) is associated with a feature representation, which induces a distance function \(\kappa(i,j)\) between items. From this distance, a similarity matrix \(M\) is defined as $M(i,j) = e^{-\kappa(i,j)}$.
Given a recommendation set \(Q \subseteq V\), the utility function is:
\begin{align}
f(Q) = \sum_{i \in V} \max_{j \in Q} M(i,j).
\end{align}
This is a monotone, submodular function, and measures how well the selected items
represent the full item set.

We represent the linear cost function $c(Q) = \sum_{i \in Q} w_i$,
where \(w_i\) denotes the Euclidean distance of restaurant \(i\) from the city center. 
The diameter cost follows from the graph \(G=(V,E)\) with nodes characterized by their geographical coordinates and edge weights \(d(i,j)\) corresponding to the geographical distances.

\spara{Influence Maximization.}
This formulation captures the trade-off between influence spread and
resource expenditure in viral marketing and social network intervention tasks.
Consider a social network \(G=(V,E)\) and an information propagation model \(P\). The goal of influence maximization is to pick a set of seed nodes \(Q\subseteq V\) such that the spread of influence (e.g., an idea or product) to other nodes in the social network is maximized.

Let \(f\) denote the expected number of nodes influenced using an Independent Cascade or Linear Threshold model for $P$. 
Given a seed set \(Q\subseteq V\), it is known that $f(Q)$ is monotone, submodular~\cite{kempe2003maximizing}. 
For linear costs, we associate a cost \(w_i\) with each node \(i \in V\), modeling the operational cost of targeting that individual (e.g., incentives or advertising spend), yielding
\(c(Q) = \sum_{i \in Q} w_i\).
To model coordination constraints among seeds, we define a graph
\(G=(V,E)\) where edge weights \(d(i,j)\) encode communication or geographical
distances between users, and instantiate the diameter cost as
\(c(Q) = \max_{i,j \in Q} d(i,j)\), encouraging well-coordinated seed sets.

%% file: 5algorithms.tex
\section{Computing Approximate Pareto Frontiers}
\label{sec:algorithm}
Papadimitriou and Yannakakis~\cite{papadimitriou2000approximability} showed that polynomial-size \emph{approximate Pareto frontiers} can be computed via discretization of the objective space. Their framework relies on
the ability to solve constrained single-objective subproblems.
In our setting, this does not hold: constrained maximization of $f$ (resp.\ minimization of $c$) subject to a constraint on $c$ (resp.\ $f$) is $NP$-hard.
We get around this computational obstacle, and present algorithms with approximation guarantees for different instantiations of {\paretosubmodcost}; proofs are deferred to an extended version of the work.

\subsection{Algorithms for {\paretosubmodknapsack}}
\label{sec:alg-linear}
Let us start by considering two special cases of {\paretosubmodknapsack}  in order to gain some intuition for the general algorithm.

\spara{The {\paretosubmodcardinality} Problem}. This is a variant of
{\paretosubmodknapsack}, where the items have uniform cost {\ie}, for each $i \in V$, $w_i = 1$; note that the {\cardinality} cost function can only take integer values in $\{0,\dots,n\}$, and thus the Pareto frontier has size at most $n$; {\ie}, polynomial in the size of the input.

We find an approximate Pareto frontier for this problem using the {\cgreedy} algorithm (see Alg.~\ref{alg:c-greedy}).
The algorithm exploits the discreteness of the cardinality cost, and constructs a set of cost thresholds
$\mathcal{B}=\{1,2,\ldots,n\}$; for each threshold $B\in\mathcal{B}$,
invokes the greedy procedure $\greedy(f,c_k,B, 0)$ to approximately solve
\[
\max_{S \subseteq \groundset} \; f(S) \quad \text{s.t.} \quad |S| \le B .
\]
Each threshold yields a candidate solution that approximates the optimal utility
achievable at the corresponding cardinality level.

\begin{algorithm}
\caption{\cgreedy$(f,c,\mathcal{B},\tau)$}
\label{alg:c-greedy}
\begin{algorithmic}[1]
\Statex \textbf{Input:} Ground set $\groundset$, monotone submodular function $f$, cost function $c$, set of cost thresholds $\mathcal{B}$, seed size $\tau$
\State $\mathcal{P} \gets \emptyset$
\ForAll{cost thresholds $B \in \mathcal{B}$}
    \State $S_B \gets$ \greedy$(f,c,\infty,B,\tau)$
    \State $\mathcal{P} \gets \mathcal{P} \cup \{(f(S_B), c(S_B))\}$
\EndFor
\State \Return \textsc{Pareto-Prune}$(\mathcal{P})$
\end{algorithmic}
\end{algorithm}

The algorithm collects all such  solutions and applies Pareto pruning;
\textsc{Pareto-Prune} removes any solution $S$ for which there
exists another solution $S'$ with
$f(S') \ge f(S)$ and $c(S') \le c(S)$, with at least one inequality strict.
By the standard approximation guarantee associated with {\greedy}~\cite{nemhauser1978analysis,vazirani2001approximation} we obtain Proposition~\ref{proposition:c-greedy-cardinality}.

\begin{proposition}\label{proposition:c-greedy-cardinality}
Given cardinality thresholds $\mathcal{B} = \{1,\ldots,n\}$ and $\tau \ge 0$, the {\cgreedy} algorithm returns an $\left(1-\frac{1}{e},1\right)$--approximate Pareto frontier $\mathcal{P}$ of size $\bigO(n)$ for {\paretosubmodcardinality}. 
\end{proposition}

In practice, the set of solutions in $\mathcal{P}$
can be obtained more efficiently.
A single run of the greedy algorithm from the empty set produces the greedy
solution for every cardinality threshold simultaneously: after the $i$-th
iteration, the current solution has cardinality $B_i$ and coincides with the
solution returned by $\greedy(f,c_k,B_i)$.
Thus, \cgreedy\ can be implemented using a single \greedy run without affecting
its guarantees.

\spara{{\paretosubmodknapsack} with Polynomial Number of f Values.}
We next consider a complementary special case of  {\paretosubmodknapsack}
where the submodular utility function $f$ takes polynomially many values.
This setting arises naturally in several applications.
For instance, in the team formation objective defined in
Eq.~\eqref{eqn:cov-objective}, $f$ takes values in
$\{0,\tfrac{1}{|T|},\ldots,1\}$, where $|T|$ is the (polynomial)
number of skills in the task.
Similarly, for the influence maximization objective, $f$ takes integer values in
$\{0,1,\ldots,|\groundset|\}$.

For such cases, the Pareto frontier can be approximated efficiently using the  \fgreedy\ algorithm, which enumerates candidate solutions
corresponding to different target values of the utility.
Let $\mathcal{K}$ denote the set of attainable values of $f$.
For each target $K \in \mathcal{K}$, \fgreedy\ invokes the greedy procedure
$\greedy(f,c,K,0)$ to approximately solve the submodular cover problem
$$\min_{S \subseteq \groundset} \; c(S)
\quad \text{s.t.} \quad f(S) \ge K.$$
From the standard approximation guarantee of the greedy
set cover algorithm~\cite{vazirani2001approximation}, this problem admits an
$\bigO(\log n)$-approximation in polynomial time.

\begin{algorithm}
\caption{\fgreedy$(f,c,\mathcal{K},\tau)$}
\label{alg:f-greedy}
\begin{algorithmic}[1]
\Statex \textbf{Input:} Ground set $\groundset$, monotone submodular function $f$, cost function $c$, set of utility targets $\mathcal{K}$, seed size $\tau$
\State $\mathcal{P} \gets \emptyset$
\ForAll{targets $K \in \mathcal{K}$}
    \State $S_K \gets$ \greedy$(f,c,K,\infty,\tau)$
    \State $\mathcal{P} \gets \mathcal{P} \cup \{(f(S_K), c(S_K))\}$
\EndFor
\State \Return \textsc{Pareto-Prune}$(\mathcal{P})$
\end{algorithmic}
\end{algorithm}

Each target value yields a candidate solution that approximates the minimum cost
required to achieve utility at least $K$.
The algorithm collects all such candidates and applies Pareto pruning to retain
only the best tradeoffs.

\begin{proposition}\label{proposition:f-greedy-coverage}
Assuming $f$ takes polynomially many distinct values and letting $\mathcal{K}$
denote this value set, and $\tau \ge 0$, the {\fgreedy} algorithm
returns an $(1,\log n)$--approximate Pareto frontier $\mathcal{P}$ of polynomial
size for the {\paretosubmodknapsack} problem.
\end{proposition}

\spara{The {\fcgreedy} algorithm.}
The special cases discussed above illustrate how discreteness in either the
cost or the objective can be exploited to construct approximate Pareto
frontiers.
For general {\paretosubmodknapsack}, however, neither the
submodular utility function $f$ nor the linear cost function
$c_\ell(S)=\sum_{i\in S} w_i$ have a polynomially-bounded range.
As a result, the Pareto frontier may be exponentially large, and the techniques
used for \cgreedy\ and \fgreedy\ no longer apply directly.

To address this setting, we 
discretize the cost and utility axes 
by picking a polynomial number of utility and cost values (which define a grid).
We then solve the corresponding constrained-optimization problems at each grid point.
This yields two complementary approximate Pareto sets: one obtained by applying
\cgreedy\ on the cost, and one obtained by applying \fgreedy\ on the utility.

\begin{algorithm}
\caption{\fcgreedy$(f,c,\mathcal{K},\mathcal{B},\tau)$}
\label{alg:fc-greedy}
\begin{algorithmic}[1]
\Statex \textbf{Input:} Ground set $\groundset$, submodular function $f$, cost function $c$, utility targets $\mathcal{K}$, budget thresholds $\mathcal{B}$, seed size $\tau$

\State $\mathcal{P}_c \gets$ \cgreedy$(f,c,\mathcal{B},\tau)$

\State $\mathcal{P}_f \gets$ \fgreedy$(f,c,\mathcal{K},\tau)$
\State \Return \textsc{Pareto-Prune}$(\mathcal{P}_c \cup \mathcal{P}_f)$
\end{algorithmic}
\end{algorithm}
The construction of the grids $\mathcal{K}$ and $\mathcal{B}$ directly affects
both the approximation quality and the computational complexity of
{\fcgreedy}. We propose two reasonable grid constructions.

\emph{Logarithmic-$\varepsilon$ grids.}
Following Papadimitriou and Yannakakis~\cite{papadimitriou2000approximability},
given an accuracy parameter $\varepsilon>0$, a utility function $f$ taking
values in $[f_{\min}, f_{\max}]$, and a cost function $c$ taking values in $[c_{\min}, c_{\max}]$ we define their corresponding grids:
\begin{align*}
    \mathcal{K}_{\varepsilon} =\{f_{\min},\, f_{\min}(1+\varepsilon),\, f_{\min}(1+\varepsilon)^2,\ldots, f_{\max}\}\\
    \mathcal{B}_{\varepsilon}
=\{c_{\min},\ldots ,c_{\max}(1-\varepsilon)^2,c_{\max}(1-\varepsilon),c_{\max}\}
\end{align*}
These grids have size polynomial in the input parameters and logarithmic in the
dynamic range of $f$ and $c$.

\emph{Linear-$\Delta$ grids.}
Alternatively, we consider linear grids parameterized by a step size $\Delta>0$. These are heuristic and do not provide worst-case approximation guarantees, but are often simpler to tune in practice. Specifically, we define
\[
\mathcal{K}_{\Delta}
=\{f_{\min},\, f_{\min}+\Delta,\, f_{\min}+2\Delta,\, \ldots,\, f_{\max}\},
\]
\[
\mathcal{B}_{\Delta}
=\{c_{\min},\, c_{\min}+\Delta, c_{\min}+2\Delta,\, \ldots,\, c_{\max}\}.
\]

Combining propositions~\ref{proposition:c-greedy-cardinality} and~\ref{proposition:f-greedy-coverage}, together with the results of ~\cite{papadimitriou2000approximability,feldman2023practical}, we get the following result:

\begin{proposition}\label{proposition:fc-greedy}
For the $\mathcal{K}_{\varepsilon}$ and $\mathcal{B}_{\varepsilon}$ constructed as above, and $\tau = 2$, the {\fcgreedy} algorithm
outputs an
$\left((1-\tfrac{1}{e})(1-\varepsilon),\; (1+\varepsilon)\,O(\log n)\right)
$-approximate Pareto frontier $\mathcal{P}$ of polynomial size
for the general {\paretosubmodknapsack} problem.
\end{proposition}

\paragraph{Discussion.}
The \fcgreedy\ algorithm provides a general and theoretically grounded approach
for approximating Pareto frontiers in the general
{\paretosubmodknapsack} setting when logarithmic grids
$\mathcal{K}_{\varepsilon}$ and $\mathcal{B}_{\varepsilon}$ are used.
However, the need to solve one constrained subproblem per grid point makes the
approach polynomial but often impractical on large instances.
This motivates the development of a faster algorithm that approximates the Pareto
frontier using a grid-free approach and fewer greedy runs. 

\spara{The {\paretogreedy} algorithm.}
We introduce an efficient alternative that directly enumerates candidate
tradeoffs using a small number of greedy runs.
Rather than solving separate subproblems for each budget level,
{\paretogreedy} explicitly collects these intermediate solutions (prefixes) and
applies Pareto pruning.

Inspired by prior work on submodular maximization with knapsack
constraints~\cite{sviridenko2004note,feldman2023practical}, 
{\paretogreedy} enumerates all feasible seed sets
of size at most $\tau$ (typically $\tau\in\{1,2\}$).
For each seed $S_0$, the algorithm runs {\greedy} up to a maximum budget
$B$, corresponding to the largest cost of interest on the Pareto frontier, and
records all intermediate solutions encountered. 
All recorded solutions are then Pareto-pruned to obtain an approximate Pareto
frontier.

\begin{algorithm}
\caption{{\paretogreedy}$(f,c,B,\tau)$}
\label{alg:pareto-greedy}
\begin{algorithmic}[1]
\Statex \textbf{Input:} Ground set $\groundset$, monotone submodular function $f$, cost function $c$, maximum budget $B$, seed size $\tau$
\State $\mathcal{P} \gets \emptyset$
\ForAll{seed sets $S_0 \subseteq V$ with $|S_0|\le\tau$ and $c(S_0)\le B$}
    \State $S \gets S_0$, $\mathcal{P} \gets \mathcal{P} \cup \{(f(S),c(S))\}$
    \While{there exists $u \notin S$ $c(S \cup \{u\}) \le B$}
        \State $u \in \arg\max_{v \notin S}
        \dfrac{f(S\cup\{v\}) - f(S)}{c(\{v\})}$
        \State $S \gets S \cup \{u\}$
        \State $\mathcal{P} \gets \mathcal{P} \cup \{(f(S),c(S))\}$
    \EndWhile
\EndFor
\State \Return \textsc{Pareto-Prune}$(\mathcal{P})$
\end{algorithmic}
\end{algorithm}

\begin{proposition}\label{proposition:pareto-greedy}
{\paretogreedy} returns a
$\bigl(1-(1-\beta)e^{-\gamma},\,1\bigr)$--approximate Pareto frontier $\mathcal{P}$
for {\paretosubmodknapsack}.
Specifically, for every Pareto-optimal solution $S^\ast$, $\mathcal{P}$ contains a
solution $S$:
\[
f(S) \;\ge\; \bigl(1-(1-\beta)e^{-\gamma}\bigr) f(S^\ast),
\qquad
c_\ell(S) \;\le\; c_\ell(S^\ast),
\]
where $\beta\in(0,1]$ is the maximum fraction of $f(S^\ast)$ captured by a seed
$S_0 \subseteq S^\ast$, and $\gamma\in(0,1]$ is the fraction of the remaining
budget $c_\ell(S^\ast)-c_\ell(S_0)$ attained by $S$.
\end{proposition}

{\paretogreedy} runs in $O(n^{\tau+2})$ time, dominated by $O(n^\tau)$ greedy runs,
each consisting of $O(n)$ iterations with $O(n)$ marginal evaluations.
Pareto pruning is asymptotically negligible.
Unlike grid-based methods, {\paretogreedy} directly enumerates feasible tradeoffs
under a fixed budget, making it significantly more practical.

\subsection{Algorithm for {\paretosubmoddiam}}
\label{sec:alg-diameter}

In the {\paretosubmoddiam} problem, the cost of a
solution is defined by its diameter in a weighted graph that defines a distance metric between 
its nodes.
Although the diameter is not a linear function, it admits a strong
geometric characterization: any set of diameter at most $B$ is contained in a
metric ball of radius $B$ centered at some vertex.
This property allows us to enumerate all cost levels implicitly,
without an explicit discretization.

\cgreedydiameter\ (Alg.~\ref{alg:diameter}) 
extends \cgreedy\ to the diameter setting by
replacing numeric cost thresholds with metric balls.
For each vertex $v\in V$, the algorithm orders all vertices by increasing
distance to $v$ and considers the resulting nested family of balls
\[
B(v,r)=\{u\in \groundset : d(v,u)\le r\},
\]
as $r$ ranges over all distinct distances.
Each such ball induces a feasible solution with diameter at most $2r$.
All candidate solutions obtained in this manner are evaluated and subsequently
Pareto-pruned to produce the approximate Pareto frontier.

\begin{algorithm}
\caption{\cgreedydiameter$(f,d(\cdot,\cdot))$}
\label{alg:diameter}
\begin{algorithmic}[1]
\Statex \textbf{Input:} Ground set $\groundset$, monotone submodular function $f$, distance metric $d(\cdot,\cdot)$
\State $\mathcal{P} \gets \emptyset$
\ForAll{centers $v \in \groundset$}
    \State Sort vertices $u\in \groundset$ by increasing $d(v,u)$
    \State $S \gets \emptyset$
    \ForAll{vertices $u$ in sorted order}
        \State $S \gets S \cup \{u\}$
        \State $\mathcal{P} \gets \mathcal{P} \cup \{(f(S),c(S))\}$
    \EndFor
\EndFor
\State \Return \textsc{Pareto-Prune}$(\mathcal{P})$
\end{algorithmic}
\end{algorithm}

\begin{proposition}\label{proposition:diameter-greedy}
The {\cgreedydiameter} algorithm returns a $(1,\,2)$--approximate Pareto frontier of size $\bigO(n^2)$ for {\paretosubmoddiam}.
\end{proposition}


For each center $v$, the algorithm sorts $n$ vertices and performs a linear scan,
yielding total running time $O(n^2\log n)$.
The number of candidate solutions is polynomial, and pruning is dominated by the
enumeration phase.
The diameter cost admits significantly more structure than general graph-based
cost functions.
Extending similar guarantees to tree or pairwise-distance costs remains open.

\subsection{Algorithm to Summarize the Pareto Frontier}
\label{sec:alg-representative}

The algorithms 
discussed in Sections~\ref{sec:alg-linear} and ~\ref{sec:alg-diameter} all return $(\alpha_1,\alpha_2)$--approximate Pareto frontiers of polynomial size, however in practice these frontiers can still have a large number of solutions. 
To address this,
we introduce the {\paretosummary} algorithm that computes a very small but representative set of points which collectively \textit{summarize} the Pareto frontier.



{\paretosummary} 
is an adaptive search method that computes a representative set of Pareto points, $\mathcal{P}_{\delta}$ that summarize the Pareto frontier, where $\delta \in (0,1)$ is a tunable tolerance parameter.
We motivate the algorithm with the special case of {\paretosubmodcardinality}, i.e. {\paretosubmodknapsack} where every item has unit cost.

We define $g(B) = \max_{c(S) \le B} f(S)$ and make the key insight that $g(B)/B$ is non-increasing as a function of the budget $B$. 
Additionally, given a budget interval $[B_\ell, B_r]$, the linear interpolation $g_L$ between the endpoints
is a convex combination of $g$ evaluated at the endpoints, i.e., for all $B \in [B_\ell, B_r]$, we have $g_L(B) = \lambda g(B_\ell) + (1-\lambda)g(B_r)$ for some $\lambda \in [0,1]$. 

{\paretosummary} constructs a representative Pareto set $\mathcal{P}_{\delta}$ by 
iteratively identifying (non-overlapping) intervals $[B_\ell, B_r]$ such that a single representative point with cost $B_\ell$ summarizes all Pareto points in the interval $[B_\ell, B_r]$.
The goal is to iteratively find the largest interval $[B_\ell, B_r]$ that satisfies $g_L(\tilde{B}) \ge (1-\delta) g(\tilde{B})$ at the midpoint $\tilde{B} = \frac{B_\ell + B_r}{2}$ of the interval, and store the solution corresponding to $g(B_\ell)$ as the `representative' for that interval.

To efficiently locate interval boundaries, {\paretosummary} uses exponential search: starting from the current left endpoint $B_\ell$, we repeatedly double the candidate right endpoint $B_r$
and query $\textsc{Pass}(B_\ell,B_r)$, which checks the interpolation condition $g_L(\tilde{B}) \ge (1-\delta) g(\tilde{B})$ at the midpoint $\tilde{B}$. The \textsc{ParetoPoint} subroutine is used to compute the solution $S$ corresponding to $g(B)$, and in practice could be any $(\alpha_1,\alpha_2)$--approximate algorithm.
Once exponential search finds an interval that fails the condition, we binary search between the last passing $B_{r_{\mathrm{prev}}}$ and the right endpoint $B_r$ to find the largest budget $B_r$ that satisfies $\textsc{Pass}(B_\ell,B_r)$.
Using the non-increasing property of $g(B)/B$ combined with our linear interpolation check, we have the following result.

\begin{proposition}
{\paretosummary} returns a $\big(\frac{\alpha_1 (1-\delta)}{\kappa}, \alpha_2\big)$ - approximate Pareto frontier for {\paretosubmodcardinality}, where $\kappa = \max \frac{B_r}{B_\ell}$ across all accepted budget intervals $[B_\ell, B_r]$, and $\alpha_1, \alpha_2$ correspond to the guarantee of the algorithm used in the \textsc{ParetoPoint} subroutine. 
\end{proposition}

While the non-increasing property of $g(B)/B$ is not always satisfied for {\paretosubmodknapsack} and {\paretosubmoddiam} in general, {\paretosummary} serves as a useful heuristic for many real-world applications.
In practice, {\paretosummary} returns a very small set of points, $\mathcal{P}_{\delta}$ since it avoids uniform grid enumeration and instead adaptively discovers the natural granularity of the Pareto frontier by leveraging the shape of the diminishing returns curve $g$. Geometrically, steeper regions of the Pareto frontier are summarized with smaller budget intervals, whereas less steep regions of the Pareto frontier can be summarized with larger budget intervals.

\begin{algorithm}
\caption{\paretosummary$(f, c, B_{\min}, B_{\max},\delta)$}
\label{alg:representative-summary}
\begin{algorithmic}[1]
\Statex \textbf{Input:} Ground set $\groundset$, submodular function $f$, cost function $c$, Budget range \([B_{\min},B_{\max}]\), tolerance \(\delta\)

\State \(\mathcal{P}_{\delta} \gets \emptyset\)
\State \(B_\ell \gets B_{\min}\)

\While{\(B_\ell < B_{\max}\)}
    \State \(B_r, B_{r_{\mathrm{prev}}} \gets \min\{2 B_\ell,B_{\max}\}, B_{\ell}\)

    \While{\(B_r<B_{\max}\) and \Call{Pass}{$B_\ell,B_r,\delta$}}
        \State \(B_{r_{\mathrm{prev}}}, B_r \gets B_r, \min\{2 B_r,B_{\max}\}\)
    \EndWhile

    \If{\Call{Pass}{$B_\ell,B_r, \delta$}}
        \State \(B_r'\gets B_r\)
    \Else
        \State \(B_r'\gets \textsc{BinarySearch}(B_{r_{\mathrm{prev}}},B_r)\)
    \EndIf

    \State \(S_l\gets \textsc{ParetoPoint}(B_\ell)\)
    \State Add \((S_l, [B_\ell, B_r'])\) to \(\mathcal{P}_{\delta}\)

    \State \(B_\ell\gets B_r'\)
\EndWhile

\State \Return \(\mathcal{P}_{\delta}\)
\end{algorithmic}
\end{algorithm}

%% file: 6experiments.tex
\section{Experiments}
\label{sec:experiments}
We examine the empirical efficacy of our algorithms on real-world instantiations of {\paretosubmodcost}. 
All algorithms are implemented in Python and executed in a single-process setting on a 64-bit Apple M1 MacBook Pro with 16~GB RAM.
To aid reproducibility, we report all experimental parameters and make our code available online\footnote{\url{https://github.com/kvombatkere/Pareto-Teams}}.

\spara{Datasets.}\label{sec:datasets}
We evaluate our methods on widely used real-world datasets spanning three application domains: team formation ({\freelancer, \bibsonomy, \imdbone, \imdbtwo}), recommender systems ({\yelpphoenix, \yelpvegas}), and influence maximization ({\NetPHY, \NetHEPT}).
For each domain, we apply application-specific preprocessing to construct the ground set of items $\groundset$, and define $f$ and $c$ as outlined in Section~\ref{sec:applications}.
Table~\ref{tab:dataset-summary} has summary statistics of the datasets, and we detail descriptions and pre-processing steps in an extended version. 

\begin{table}
\centering
\resizebox{\linewidth}{!}{
\begin{tabular}{l c cc cc cc}
\toprule
\textbf{Dataset }
& $\lvert \widehat{\groundset} \rvert$
& \multicolumn{2}{c}{\textbf{\cardinality}}
& \multicolumn{2}{c}{\textbf{\linear}}
& \multicolumn{2}{c}{\textbf{\diameter}} \\
\cmidrule(lr){3-4} \cmidrule(lr){5-6} \cmidrule(lr){7-8}
& 
& $\lvert \groundset \rvert$ & $c_k$ values
& $\lvert \groundset \rvert$ & $c_\ell$ values
& $\lvert \groundset \rvert$ & $c_d$ values \\
\midrule
\freelancer   & 50   & 50  & $[1,15]$  & 50  & $[5,100]$ & 50  & $[0,1]$ \\
\bibsonomy& 250  & 250 & $[1,15]$  & 250 & $[5,100]$ & 250 & $[0,1]$ \\
\imdbone  & 200  & 200 & $[1,15]$  & 200 & $[5,100]$ & 200 & $[0,1]$ \\
\imdbtwo  & 400  & 400 & $[1,15]$  & 400 & $[5,100]$ & 400 & $[0,1]$ \\
\yelpphoenix  & 1849 & 1000 & $[1,500]$ & 400 & $[0.1,47.4]$  & 800 & $[0,50]$ \\
\yelpvegas& 3203 & 1500 & $[1,600]$ & 600 & $[0.2,24.1]$  & 1200 & $[0,75]$ \\
\NetPHY   & 912  & 912 & $[1,200]$ & 151 & $[1,47]$  & 912 & $[0,50]$ \\
\NetHEPT  & 1673 & 1673 & $[1,200]$ & 354 & $[1,29]$  & 1673 & $[0,25]$ \\
\bottomrule
\end{tabular}
}
\caption{Summary of experimental datasets.
$\lvert \widehat{\groundset} \rvert$ denotes the full ground-set size, while
$\lvert \groundset \rvert$ denotes the number of items sampled from
$\widehat{\groundset}$ per experiment.
For each cost function, we report the range of values defined during
preprocessing.}
\label{tab:dataset-summary}
\end{table}

\spara{Metrics.} 
For each dataset, we evaluate all algorithms on 10 randomly sampled subsets of the ground set, with subset sizes 
$\lvert \groundset \rvert$ chosen per dataset as summarized in Table~\ref{tab:dataset-summary}. 
We evaluate all algorithms using the following metrics.

\emph{Frontier size.} This is a quantitative measure of how succinctly the algorithm constructs a \emph{representative} set of solutions. Note that all our algorithms return a polynomial-number of representative solutions (Propositions~\ref{proposition:c-greedy-cardinality}-\ref{proposition:diameter-greedy}). In many instances they may return larger number of solutions that other heuristics; this number is still polynomial in size and the difference indicates that the heuristics simply leave a large part of the Pareto frontier unexplored.

\emph{Frontier shape.} To visualize results across multiple samples, we employ an averaging heuristic; for each algorithm and sample, we linearly interpolate the computed frontier onto a fixed cost grid, producing aligned utility vectors of equal length. We then compute the mean and standard deviation of the utility at each grid point across samples, and plot the mean frontier with a shaded standard-deviation band. This interpolation-based aggregation provides a consistent visualization across heterogeneous sample frontiers and is intended as a visualization heuristic. It should be interpreted as the ``average Pareto curve" across samples.
We place markers along the mean curve in proportion to the average frontier size, distributing them evenly across the cost grid to reflect relative frontier density. Since {\paretosummary} selects a small sample of representative Pareto points, we visualize the results of this algorithm on single samples and show representative examples in Figure~\ref{fig:single-run-pareto-frontiers}.
In all our plots, curves that are higher indicate better utility--cost tradeoffs since they achieve larger utility for the same cost.

\begin{figure*}
\begin{subfigure}{\textwidth}
\centering
\includegraphics[width=0.4\textwidth]
{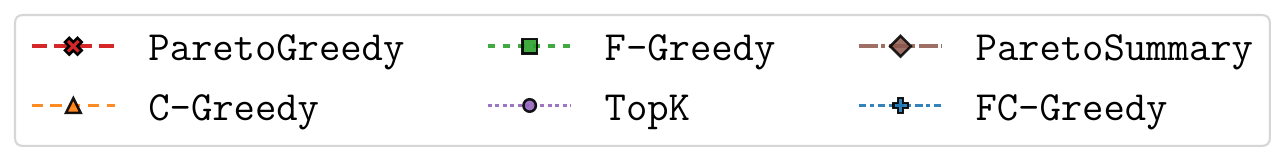}
\end{subfigure}
\centering
\begin{subfigure}{0.247\textwidth}
\includegraphics[width=\linewidth]{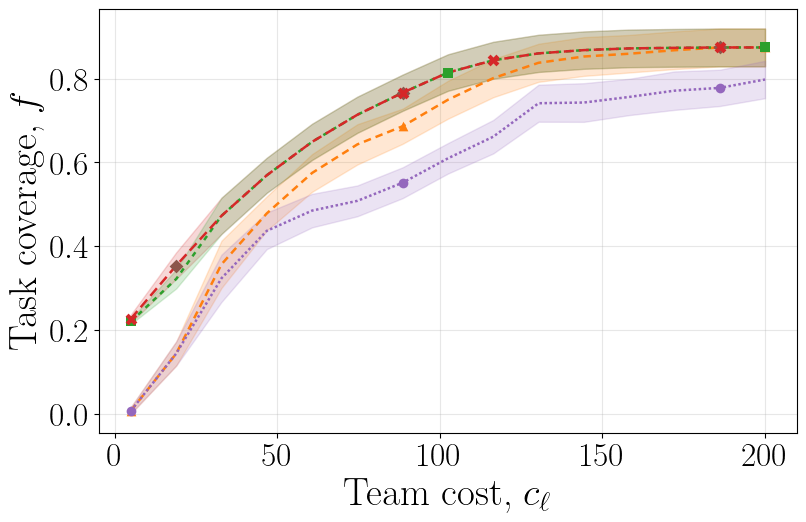}
\caption{\freelancer}
\end{subfigure}
\begin{subfigure}{0.247\textwidth}
\includegraphics[width=\linewidth]{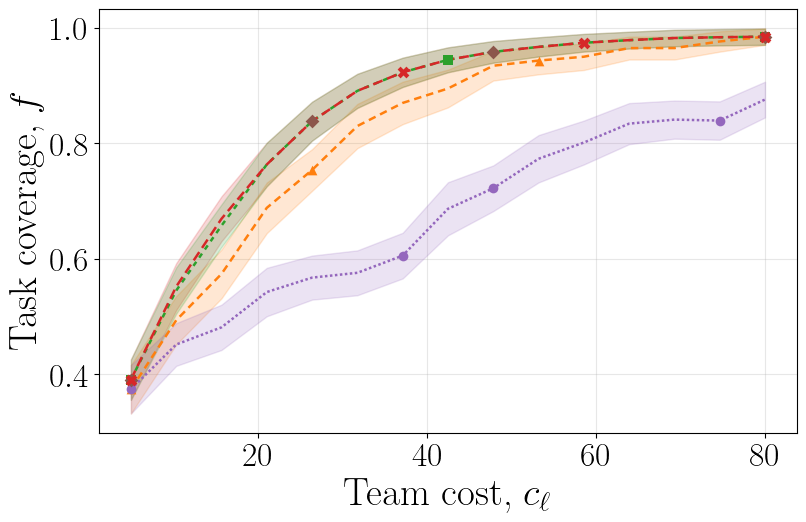}
\caption{\bibsonomy}
\end{subfigure}
\begin{subfigure}{0.247\textwidth}
\includegraphics[width=\linewidth]{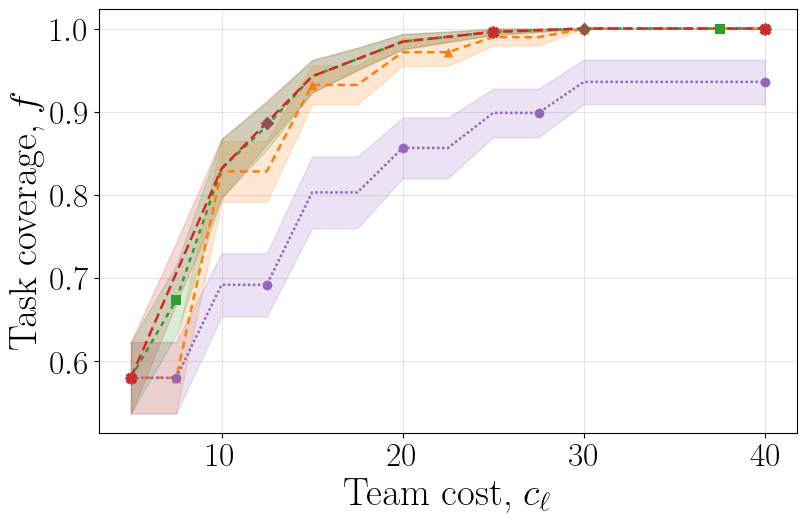}
\caption{\imdbone}
\end{subfigure}
\begin{subfigure}{0.247\textwidth}
\includegraphics[width=\linewidth]{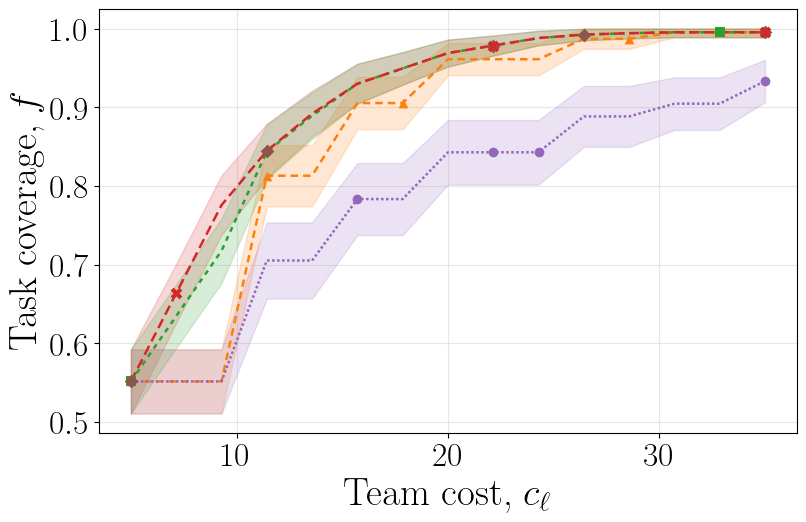}
\caption{\imdbtwo}
\end{subfigure}

\begin{subfigure}{0.247\textwidth}
\includegraphics[width=\linewidth]{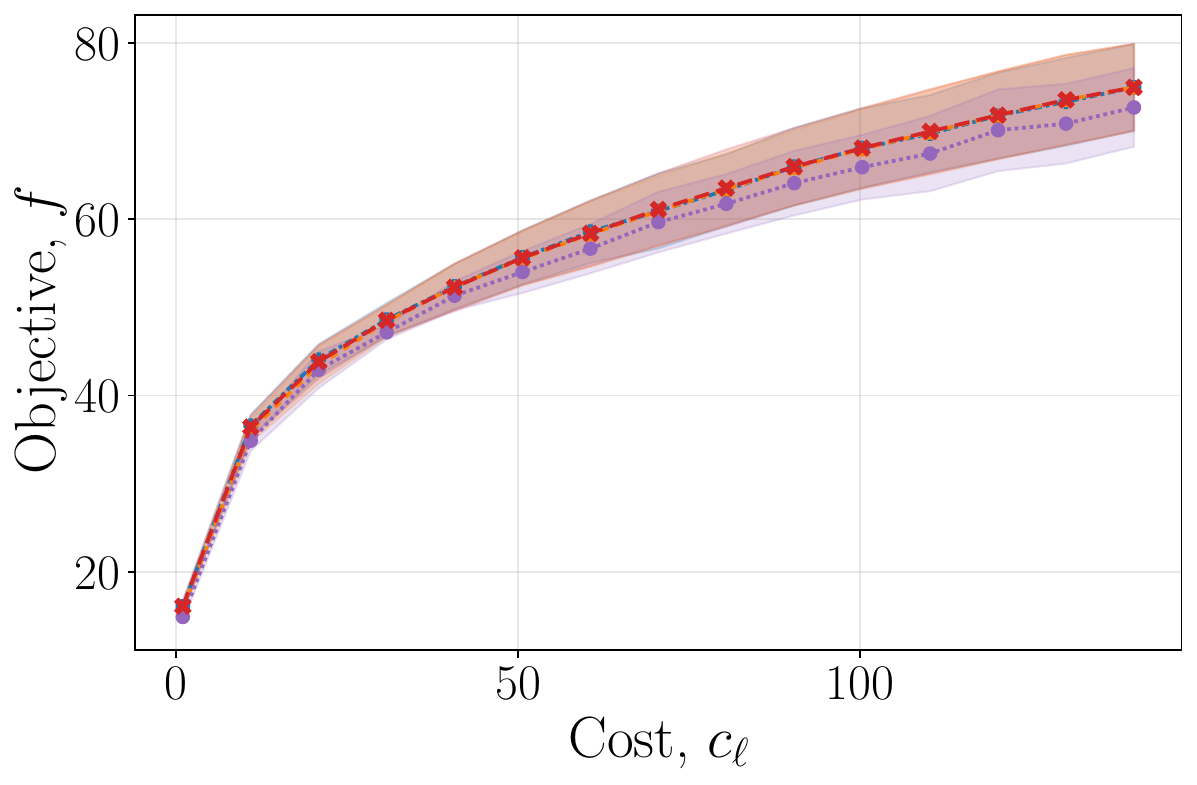}
\caption{\yelpphoenix}
\end{subfigure}
\begin{subfigure}{0.247\textwidth}
\includegraphics[width=\linewidth]{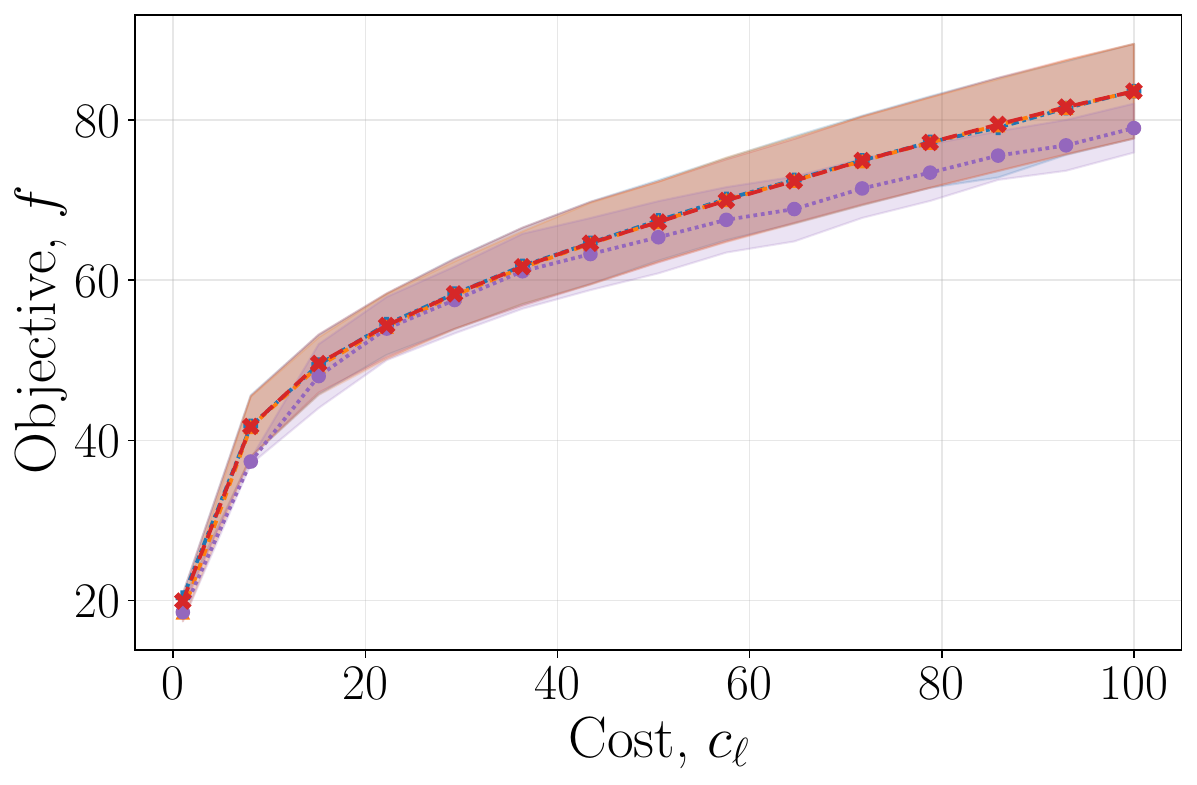}
\caption{\yelpvegas}
\end{subfigure}
\begin{subfigure}{0.247\textwidth}
\includegraphics[width=\linewidth]{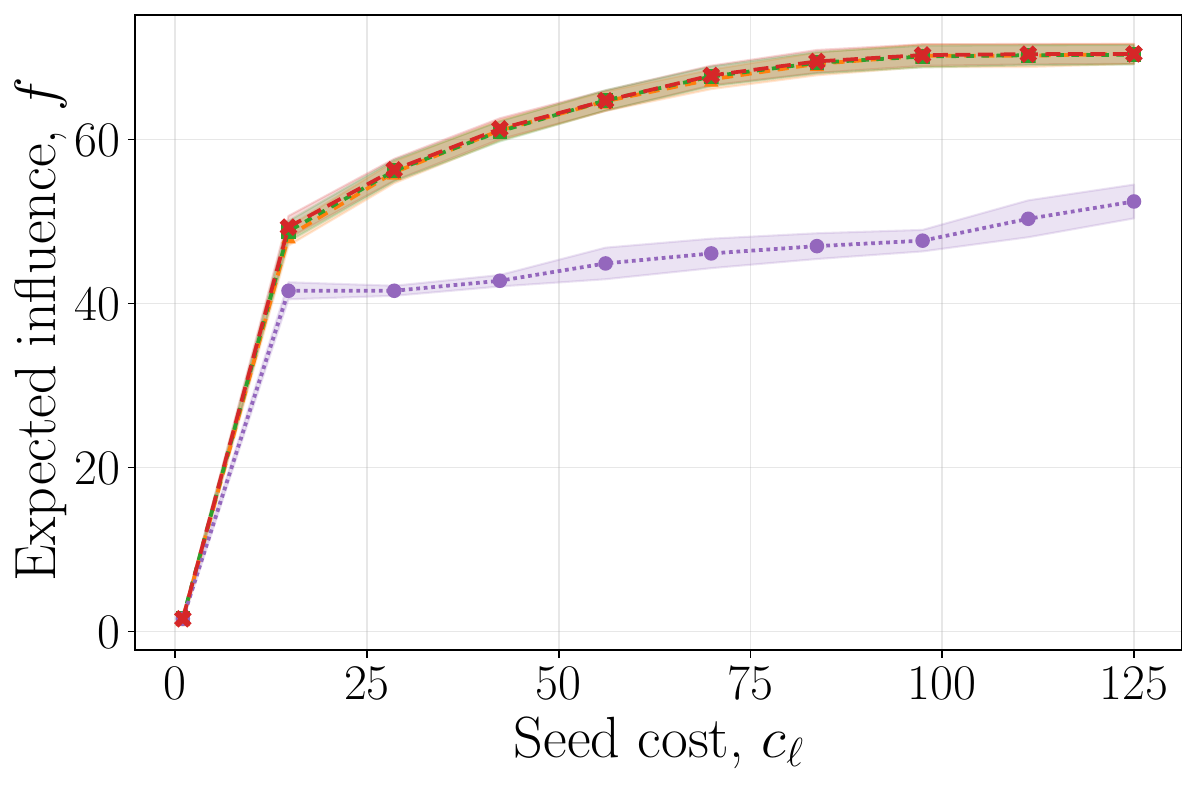}
\caption{\NetPHY}
\end{subfigure}
\begin{subfigure}{0.247\textwidth}
\includegraphics[width=\linewidth]{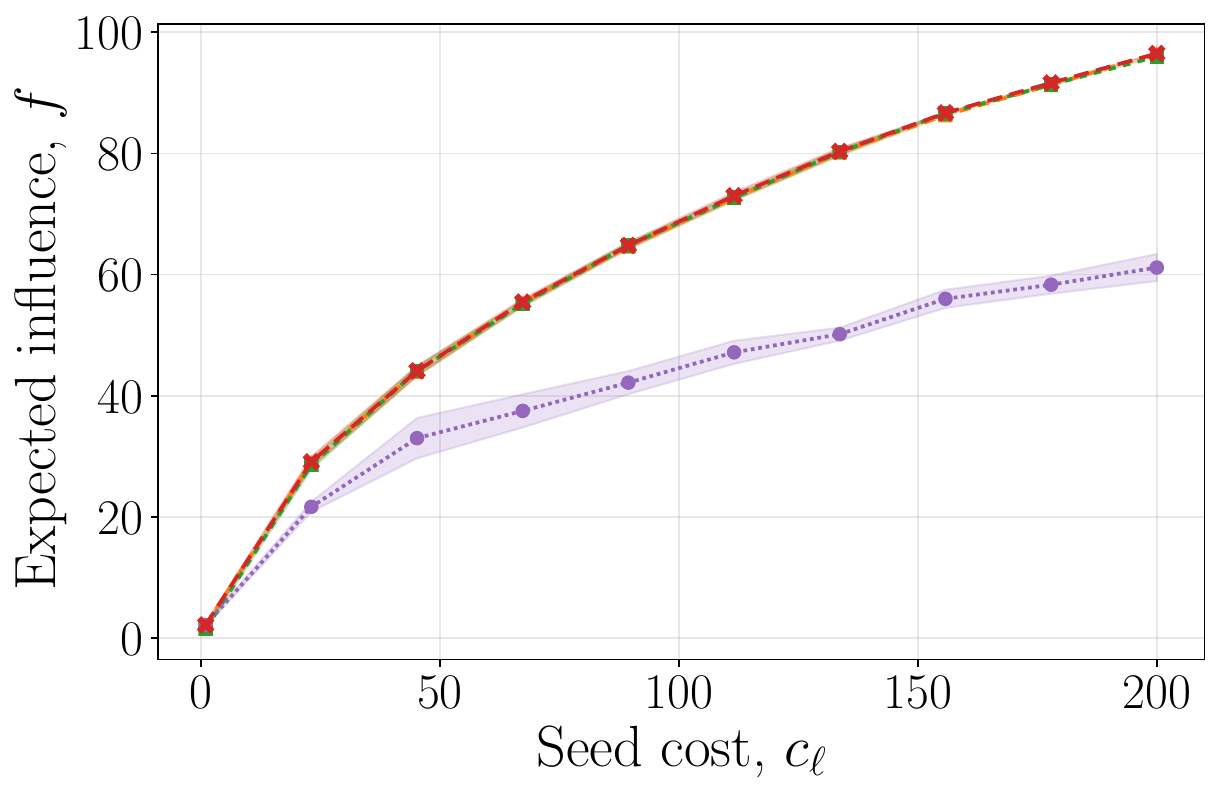}
\caption{\NetHEPT}
\end{subfigure}
\caption{Approximate Pareto frontiers averaged across multiple samples of each dataset, for all algorithms for the
{\paretosubmodknapsack} problem.
{\fcgreedy} is evaluated using logarithmic
$\varepsilon$-grids $\mathcal{K}_{0.1}$ and $\mathcal{B}_{0.1}$, while {\cgreedy} and {\baselinetopk} are baseline methods executed using a linear cost
grid $\mathcal{B}_{\Delta}$.}
\label{fig:knapsack-pareto-frontiers}
\end{figure*}

\subsection{Evaluation for \paretosubmodknapsack}
\label{sec:results-linear}
We evaluate the empirical performance and running time of the algorithms for the {\paretosubmodknapsack} problem. Due to space constraints, we provide detailed results for the special case of {\paretosubmodcardinality} in an extended version of the work.

We evaluate \fcgreedy\ using logarithmic $\varepsilon$-grids $\mathcal{K}_{0.1}$ and $\mathcal{B}_{0.1}$ and apply it only to the Yelp datasets since the  utility
$f(Q)=\sum_{i\in V}\max_{j\in Q} M(i,j)$ can admit exponentially many distinct values.
In contrast, in the influence maximization and team formation settings, the
submodular objective $f$ has a polynomially bounded value range, allowing
\fgreedy\ to be applied directly without utility discretization. We use $\delta=0.1$ for {\paretosummary}, and a fixed seed size of $\tau=1$ across all experiments to ensure consistency.

\spara{Baselines.}\label{sec:reference-linear}
\emph{\cgreedy} serves as our primary baseline. We adapt the budgeted maximization algorithm of Feldman \etal \cite{feldman2023practical} using a linear cost
grid $\mathcal{B}_{\Delta}$ to generate a Pareto frontier.
Additionally, we include two heuristics to contextualize performance. As these methods do not directly produce an explicit Pareto frontier, we execute each over a discrete set of cost thresholds ($\mathcal{B}_{\Delta}$) and retain only the resulting non-dominated solutions.

\emph{\baselinetopk} selects elements with the largest singleton marginal utilities;
elements are ordered by cost-scaled singleton utility
$f(\{i\})/w_i$ and added greedily until the budget 
is violated.

\emph{\baselinerandom} selects a random subset $S \subseteq \groundset$ of cardinality $|S| = k$.


\begin{figure*}
\begin{subfigure}{\textwidth}
\centering
\includegraphics[width=0.4\textwidth]{figures/knapsack/summary/knapsack_legend.pdf}
\end{subfigure}
\begin{subfigure}{0.247\textwidth}
\includegraphics[width=\linewidth]{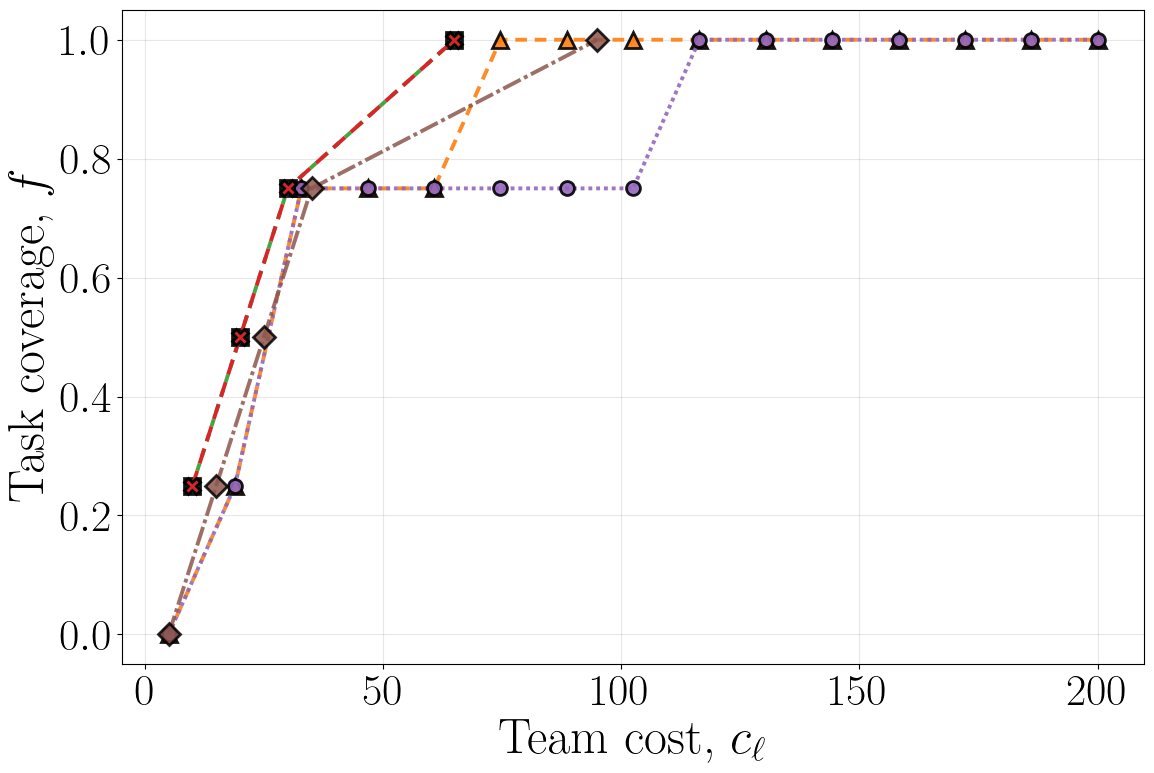}
\caption{\freelancer}
\end{subfigure}
\begin{subfigure}{0.247\textwidth}
\includegraphics[width=\linewidth]{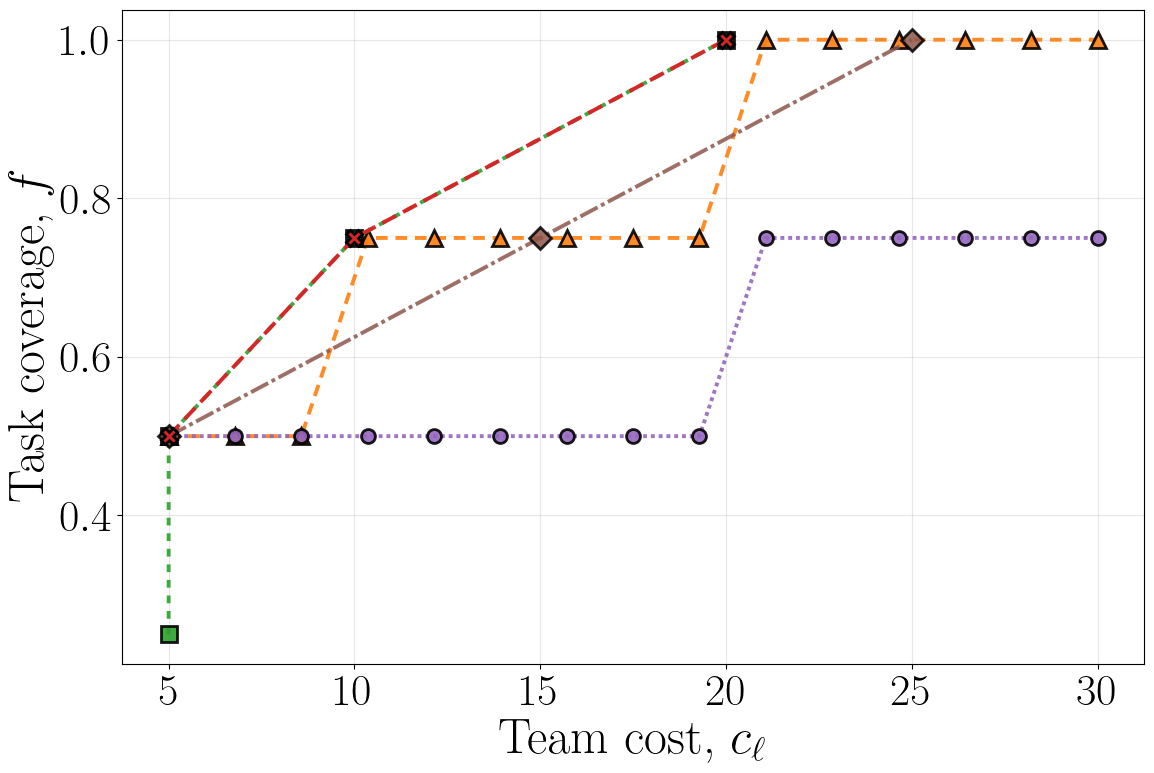}
\caption{\bibsonomy}
\end{subfigure}
\begin{subfigure}{0.247\textwidth}
\includegraphics[width=\linewidth]{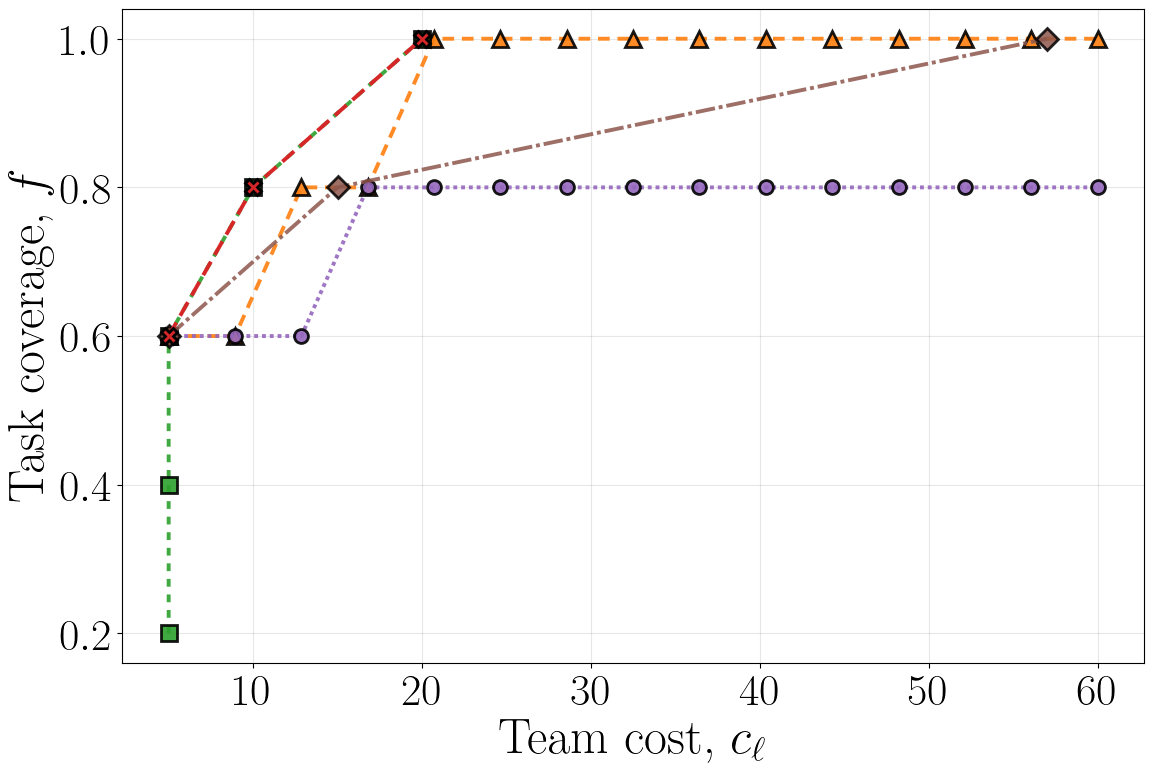}
\caption{\imdbone}
\end{subfigure}
\begin{subfigure}{0.247\textwidth}
\includegraphics[width=\linewidth]{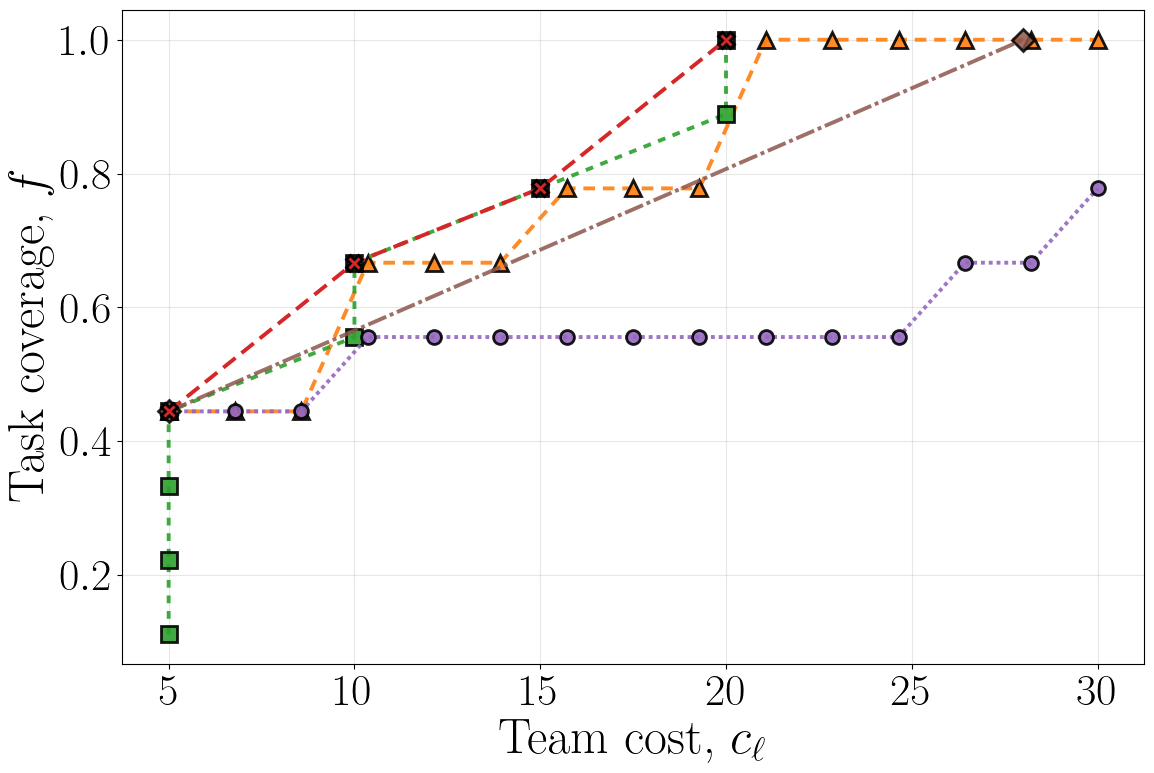}
\caption{\imdbtwo}
\end{subfigure}

\begin{subfigure}{0.247\textwidth}
\includegraphics[width=\linewidth]{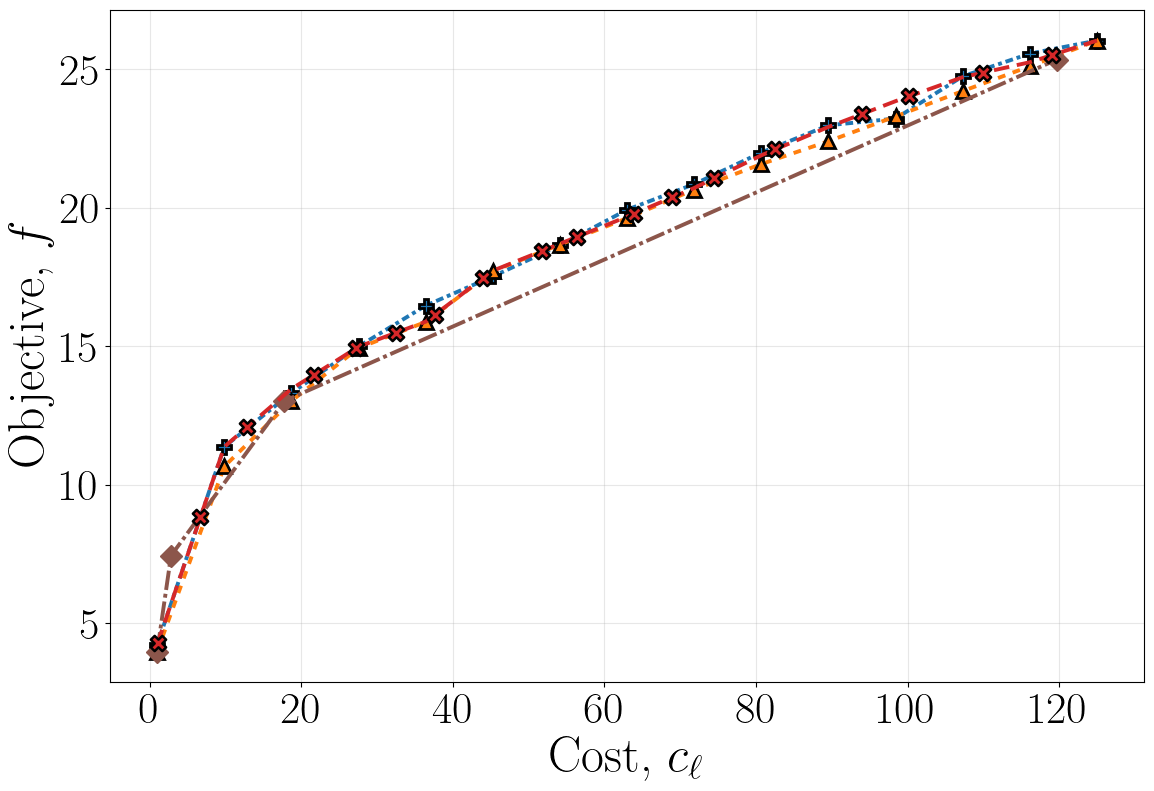}
\caption{\yelpphoenix}
\end{subfigure}
\begin{subfigure}{0.247\textwidth}
\includegraphics[width=\linewidth]{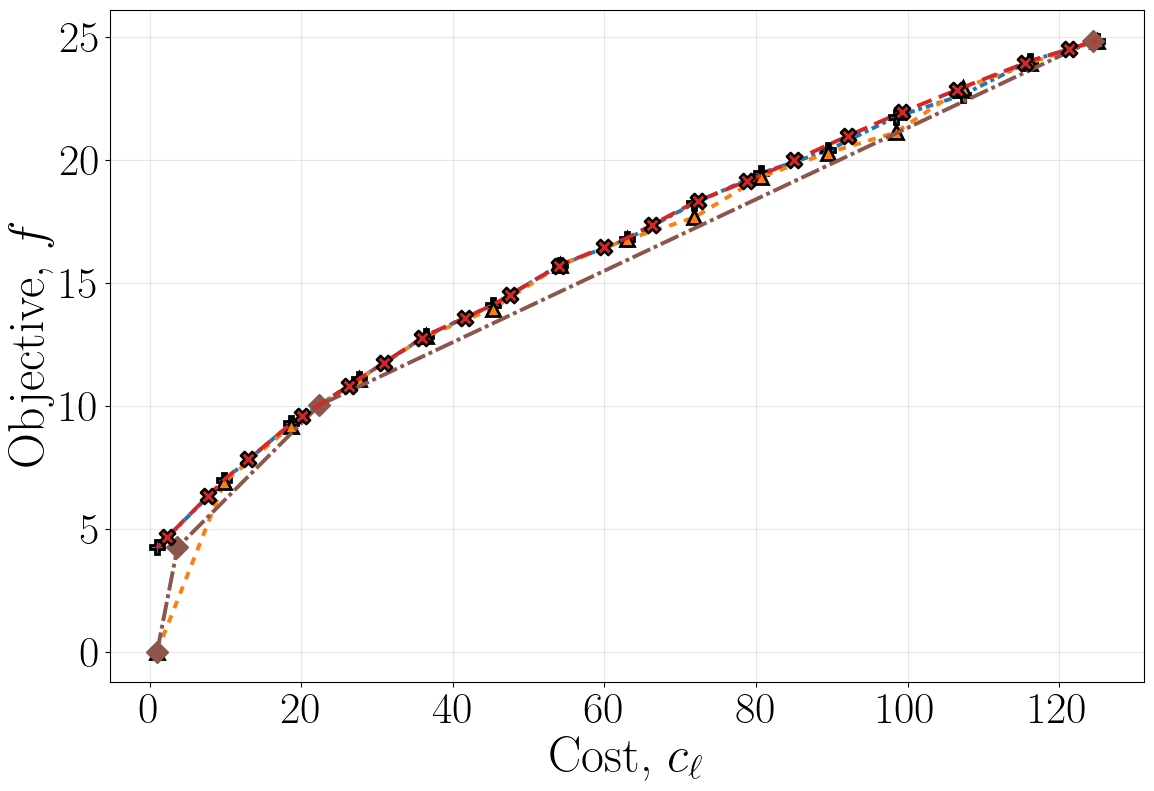}
\caption{\yelpvegas}
\end{subfigure}
\begin{subfigure}{0.247\textwidth}
\includegraphics[width=\linewidth]{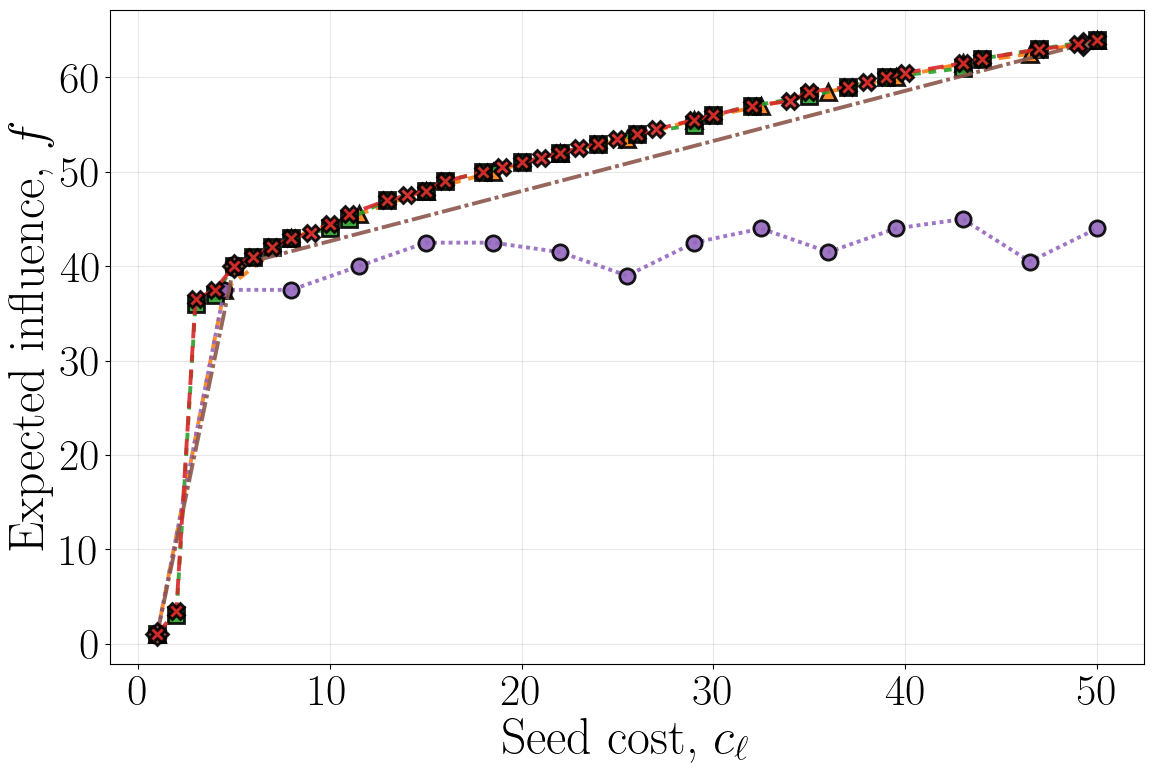}
\caption{\NetPHY}
\end{subfigure}
\begin{subfigure}{0.247\textwidth}
\includegraphics[width=\linewidth]{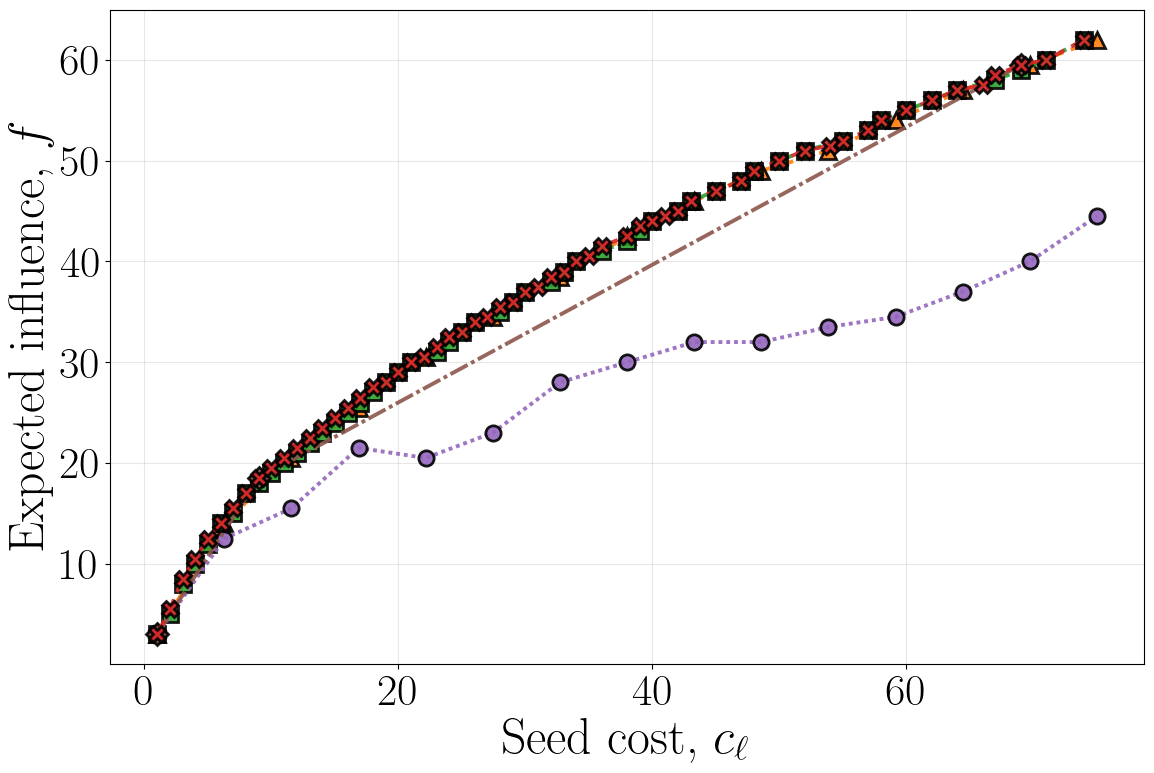}
\caption{\NetHEPT}
\end{subfigure}
\caption{Approximate Pareto frontiers computed on single samples across datasets for all algorithms for {\paretosubmodknapsack}.}
\label{fig:single-run-pareto-frontiers}
\end{figure*}

\spara{Pareto Frontier Quality.}
We first examine the qualitative structure of the utility--cost tradeoffs shown in
Figures~\ref{fig:knapsack-pareto-frontiers} and ~\ref{fig:single-run-pareto-frontiers}. \baselinerandom\ is omitted from the plots for clarity due to consistently poor performance. 
Across all datasets, \paretogreedy\ consistently traces the upper envelope of the
observed tradeoff curves, indicating strong performance across the full range of
cost values.
On team-formation datasets (\freelancer, \bibsonomy, and the \imdb\ instances),
\fgreedy\ and \paretogreedy\ achieve nearly identical frontier quality, while
grid-based methods such as \cgreedy\ exhibit noticeable gaps due to their reliance
on fixed budget discretizations.
On the \NetHEPT and \NetPHY datasets our algorithms recover
consistently better tradeoffs than \topkdegree\ with \paretogreedy consistently returning a marginally higher tradeoff frontier than \fgreedy and \cgreedy. We observe that on \yelpvegas and \yelpphoenix, all algorithms perform reasonably well in terms of their frontier shape.

We show the mean size of the approximate
Pareto frontiers produced by each algorithm in Table~\ref{tab:pareto-size}.
On smaller instances, all algorithms return succinct frontiers,
typically containing only \(3\)–\(5\) solutions.
In contrast, on larger datasets, the frontiers returned by {\fgreedy}, {\fcgreedy} and \paretogreedy\ are substantially richer.
For example, \paretogreedy\ recovers mean frontier sizes of \(177.5\) and
\(221.9\) on \yelpphoenix\ and \yelpvegas, and \(46.5\) and \(143.1\) on
\NetPHY and \NetHEPT, respectively.
These larger frontiers remain polynomial in the input size, but provide a richer representation of the underlying utility--cost tradeoff space than the baselines can capture. 
On the other hand, {\paretosummary} returns significantly smaller representative frontiers, with mean frontier sizes ranging from \(6.5\)–\(8.5\) points for these larger datasets. Additionally, Figure~\ref{fig:single-run-pareto-frontiers} demonstrates the efficacy of {\paretosummary} in representing steeper regions of the Pareto frontier with more points compared to less steep regions. In particular, {\paretosummary} adaptively identifies the `elbow' of the frontier, as well as regions which are more sensitive to a change in cost, both of which are very useful for understanding the utility--cost tradeoff.

\begin{table}
\centering
\resizebox{\linewidth}{!}{
\begin{tabular}{lcccccc}
\toprule
\textbf{Dataset}
& \textbf{\fgreedy}
& \textbf{\fcgreedy}
& \textbf{\paretogreedy}
& \cgreedy
& \baselinetopk 
& \paretosummary \\
\midrule
\freelancer& \textbf{4.1} & {\scriptsize N/A} & \textbf{3.9} & 4.7 & 4.3 & 4.1\\
\bibsonomy & \textbf{5.2} & {\scriptsize N/A}& \textbf{2.9} & 2.9 & 2.9 & 3.1\\
\imdbone   & \textbf{4.6} & {\scriptsize N/A}& \textbf{3.6} & 3.7 & 3.1 & 2.7\\
\imdbtwo   & \textbf{5.1} & {\scriptsize N/A}& \textbf{2.8} & 2.8 & 2.6 & 2.5\\
\yelpphoenix   & {\scriptsize N/A}  & \textbf{62.1} & \textbf{177.5} & 15 & 15 & 6.5\\
\yelpvegas & {\scriptsize N/A}  & \textbf{51.5} & \textbf{221.9}& 15 & 15 & 6.8\\
\NetPHY& \textbf{66.1}& {\scriptsize N/A}& \textbf{46.5} & 10 & 10 & 6.7\\
\NetHEPT   & \textbf{96.0}& {\scriptsize N/A}& \textbf{143.1}  & 10 & 10 & 8.5\\
\bottomrule
\end{tabular}}
\caption{Mean size of the approximate Pareto frontier produced by each algorithm for {\paretosubmodknapsack}.
Bold entries correspond to our algorithms, and {\cgreedy} and {\baselinetopk} are baseline methods executed using a linear cost
grid $\mathcal{B}_{\Delta}$. {\fcgreedy} is evaluated using logarithmic
$\varepsilon$-grids $\mathcal{K}_{0.1}$ and $\mathcal{B}_{0.1}$.}
\label{tab:pareto-size}
\end{table}


\begin{figure*}
\begin{subfigure}{\textwidth}
\centering
\includegraphics[width=0.4\textwidth]{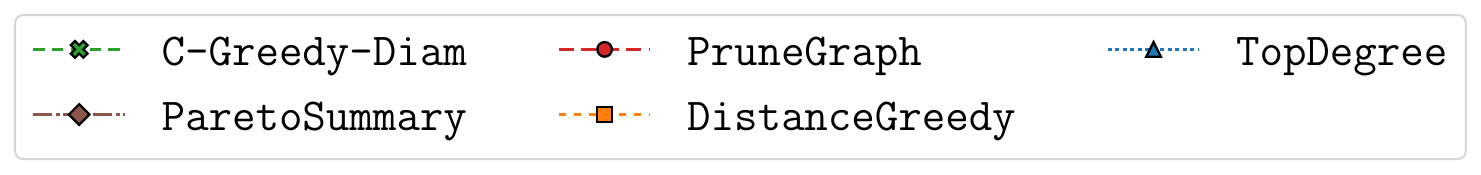}
\end{subfigure}
\begin{subfigure}{0.247\textwidth}
\includegraphics[width=\linewidth]{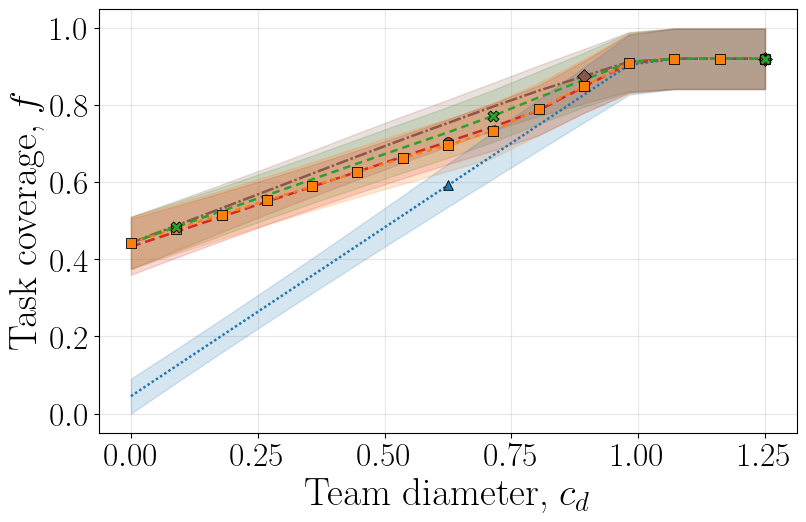}
\caption{\freelancer}
\end{subfigure}
\begin{subfigure}{0.247\textwidth}
\includegraphics[width=\linewidth]{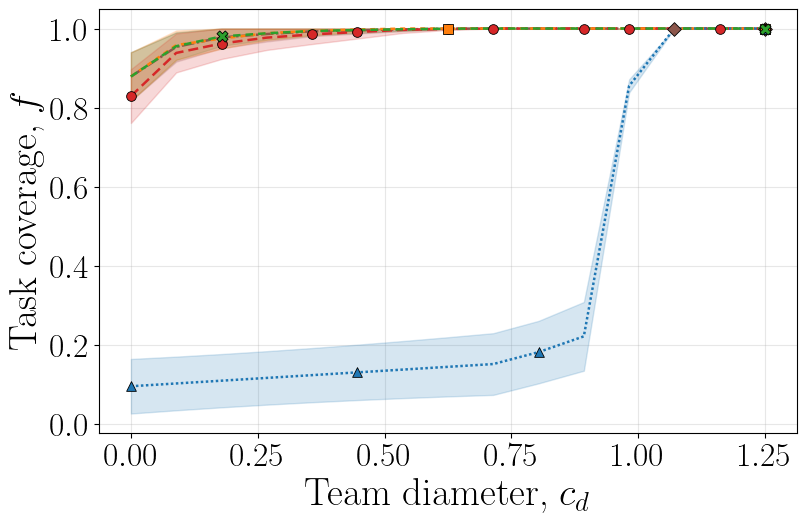}
\caption{\bibsonomy}
\end{subfigure}
\begin{subfigure}{0.247\textwidth}
\includegraphics[width=\linewidth]{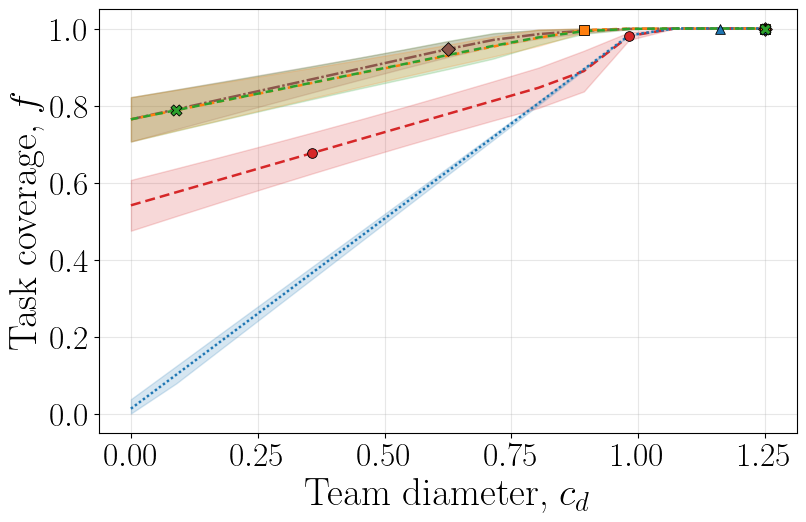}
\caption{\imdbone}
\end{subfigure}
\begin{subfigure}{0.247\textwidth}
\includegraphics[width=\linewidth]{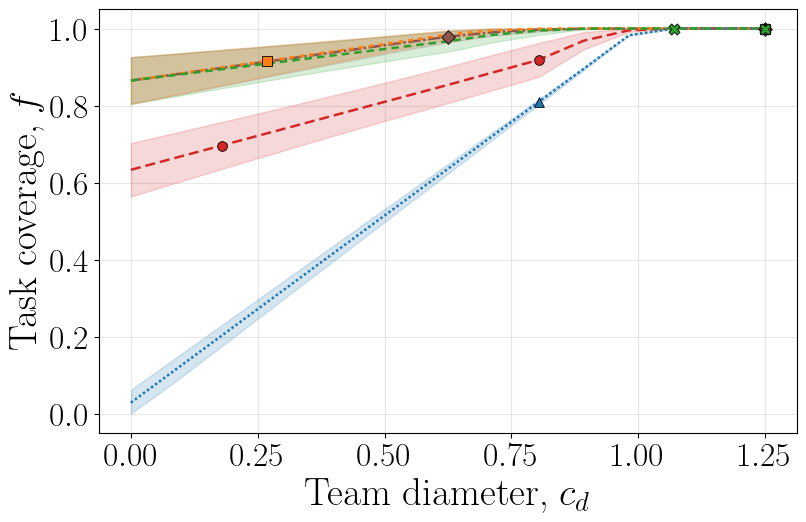}
\caption{\imdbtwo}
\end{subfigure}

\begin{subfigure}{0.247\textwidth}
\includegraphics[width=\linewidth]{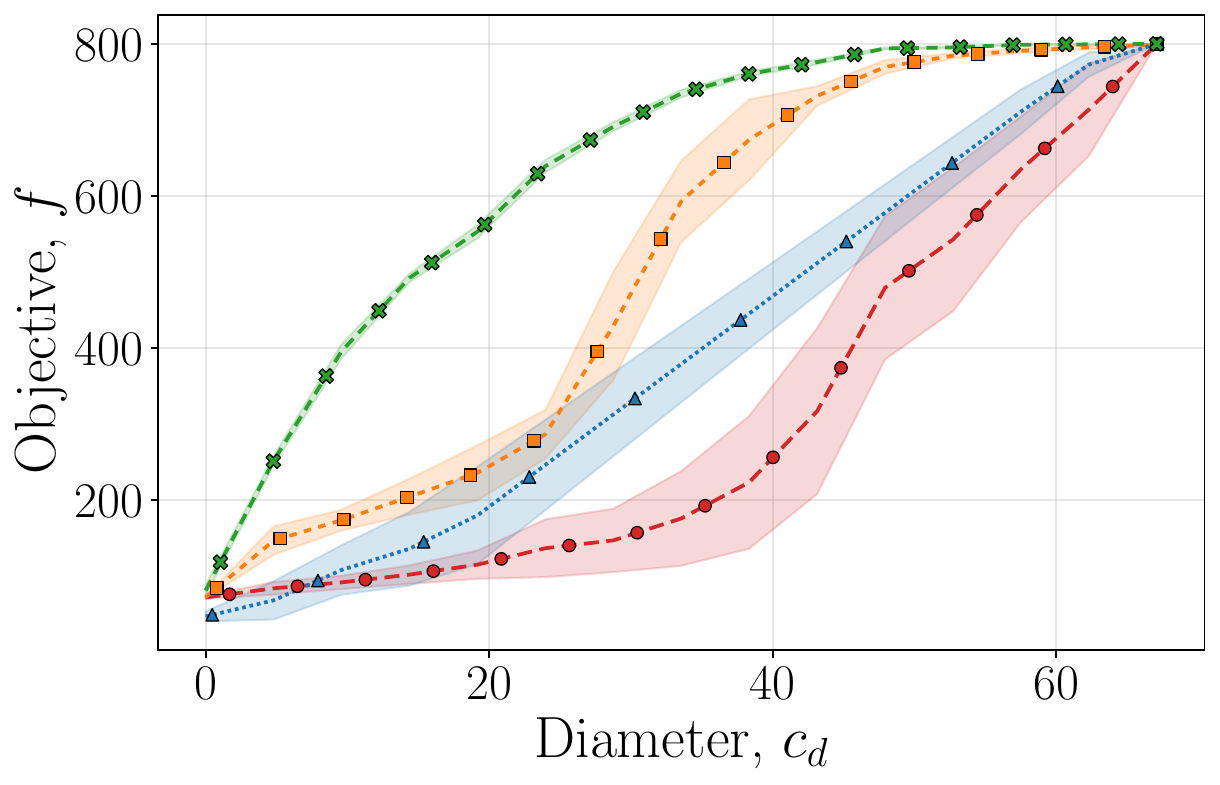}
\caption{\yelpphoenix}
\end{subfigure}
\begin{subfigure}{0.247\textwidth}
\includegraphics[width=\linewidth]{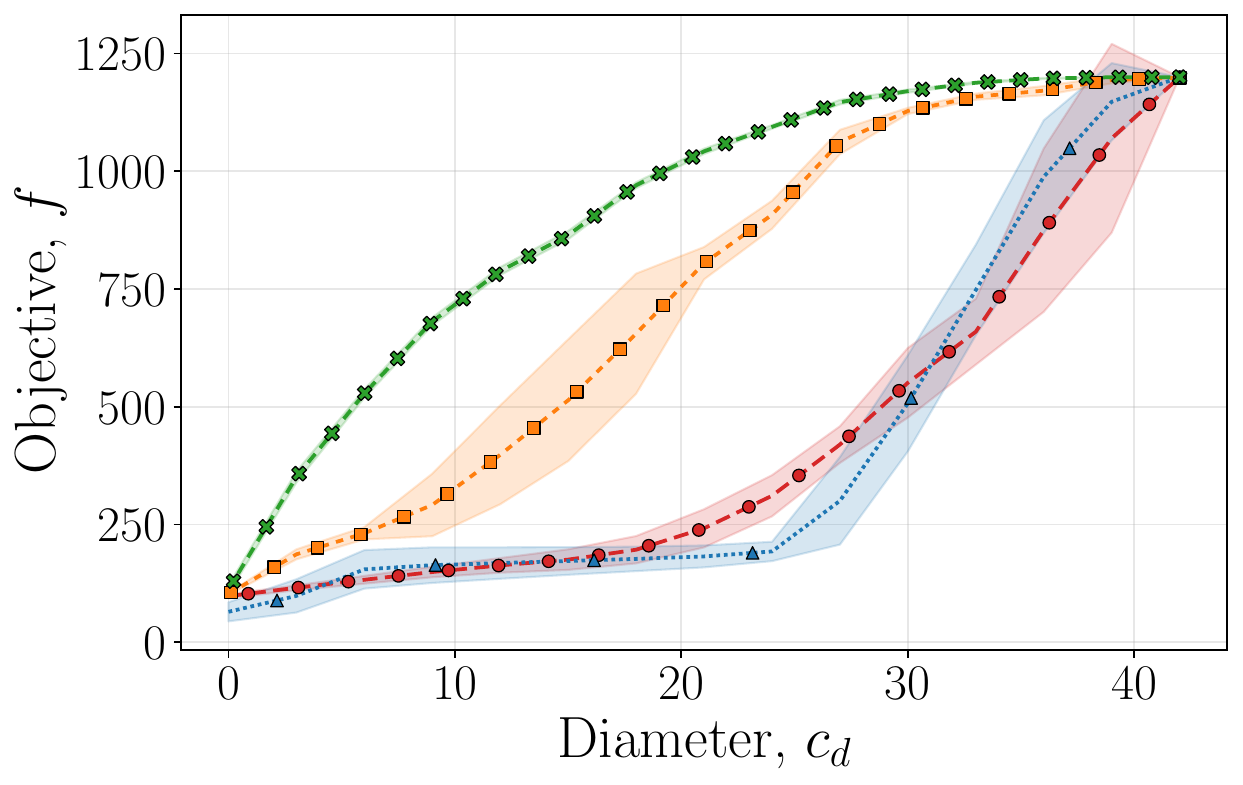}
\caption{\yelpvegas}
\end{subfigure}
\begin{subfigure}{0.247\textwidth}
\includegraphics[width=\linewidth]{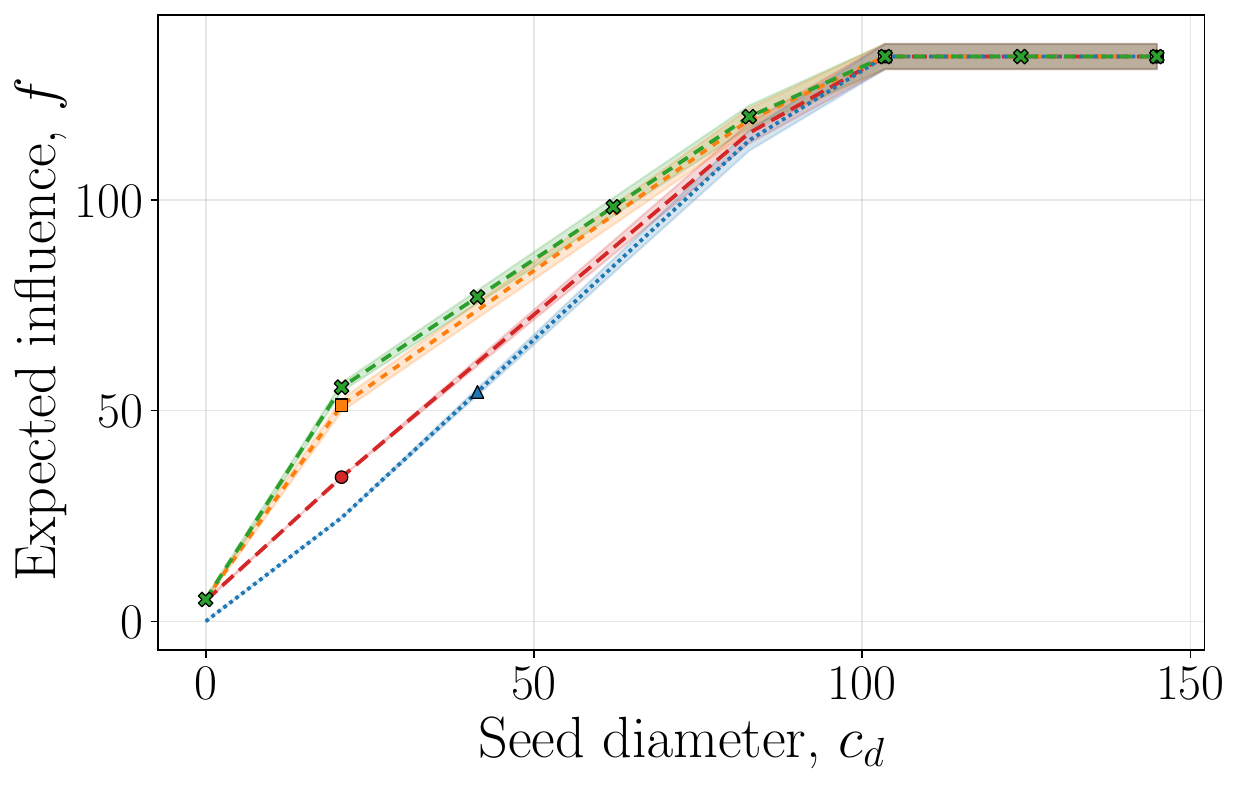}
\caption{\NetPHY}
\end{subfigure}
\begin{subfigure}{0.247\textwidth}
\includegraphics[width=\linewidth]{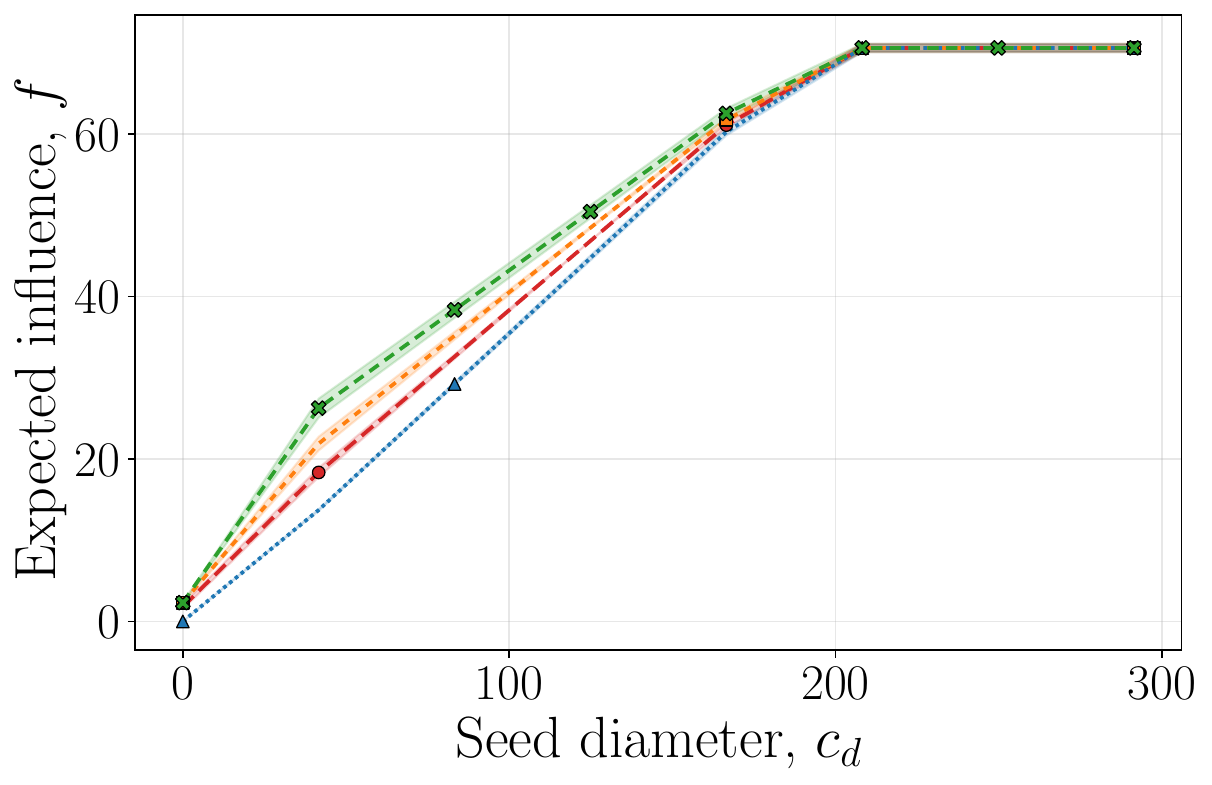}
\caption{\NetHEPT}
\end{subfigure}
\caption{Approximate Pareto frontiers for all algorithms for the {\paretosubmoddiam} problem.}
\label{fig:diameter-pareto-frontiers}
\end{figure*}

\spara{Running Time.}
Table~\ref{tab:runtime} highlights a pronounced separation between grid-based
and non-grid based approaches.
Algorithms such as \cgreedy\ and \fgreedy\ solve a large number of
budget- or utility-constrained subproblems—one per grid point—which leads to
substantial cumulative overhead as instance sizes and grid resolutions increase.
This effect is particularly evident on large datasets.
For example, on the Yelp instances, \cgreedy\ requires $1517.9$s on
\yelpphoenix\ and $3301.2$s on \yelpvegas, while \fgreedy\ incurs an even higher
cost on \NetHEPT\ ($330$s).

The \fcgreedy\ algorithm partially mitigates this cost by combining
coarser utility and budget grids, but still requires solving one constrained
subproblem per grid point.
As a result, while \fcgreedy\ improves over \cgreedy\ on the Yelp datasets
($1188.8$s vs.\ $1517.9$s on \yelpphoenix, $4527.0$s vs.\ $3301.2$s on
\yelpvegas), its running time remains dominated by repeated greedy executions.

In contrast, \paretogreedy\ achieves consistently low running times by extracting
Pareto-optimal prefixes from a small number of greedy runs.
Rather than solving separate subproblems for each budget level, it performs a
single greedy pass per seed set and records all intermediate solutions.
As a result, \paretogreedy\ completes within seconds even on the largest
instances (e.g., $125.4$s on \yelpphoenix, $394.5$s on \yelpvegas, and $7.16$s on
\NetHEPT), yielding speedups of one to two orders of magnitude over grid-based
methods. On larger datasets {\paretosummary} runs slower than \paretogreedy\ but outperforms \fcgreedy\ and \fgreedy\, since the adaptive search approach enables it to skip large sections of the frontier, on average.
While the lightweight \baselinetopk\ heuristic is fastest overall, its solutions
are substantially weaker, as shown in the Pareto-quality results.
Taken together, these findings demonstrate that \paretogreedy\ and {\paretosummary} offer a markedly
more favorable tradeoff between solution quality and computational cost for
large-scale {\paretosubmodknapsack} instances.

\begin{table}
\centering
\resizebox{\linewidth}{!}{
\begin{tabular}{lcccccc}
\toprule
\textbf{Dataset}
& \textbf{\fgreedy}
& \textbf{\fcgreedy}
& \textbf{\paretogreedy}
& \cgreedy
& \baselinetopk 
& \paretosummary \\
\midrule
\freelancer& \textbf{0.035} & {\scriptsize N/A} & \textbf{0.011} & 0.101  & 0.001 & 0.311\\
\bibsonomy & \textbf{0.090} & {\scriptsize N/A} & \textbf{0.021} & 0.117  & 0.002 & 0.840\\
\imdbone& \textbf{0.036} & {\scriptsize N/A} & \textbf{0.013} & 0.055  & 0.001 & 0.143\\
\imdbtwo& \textbf{0.046} & {\scriptsize N/A} & \textbf{0.011} & 0.072  & 0.001 & 0.163\\
\yelpphoenix  & {\scriptsize N/A}  & \textbf{1188.8 } & \textbf{125.4 } & 1517.9 & 0.062 & 653.213\\
\yelpvegas & {\scriptsize N/A}  & \textbf{4526.962} & \textbf{394.521} & 3301.184 & 0.054 & 3241.542\\
\NetPHY & \textbf{15.694} & {\scriptsize N/A} & \textbf{0.897} & 7.035  & 0.021 & 7.940\\
\NetHEPT& \textbf{329.990} & {\scriptsize N/A}& \textbf{7.157} & 41.964 & 0.051 & 202.56\\
\bottomrule
\end{tabular}}
\caption{Running time comparison (in seconds) of algorithms for
{\paretosubmodknapsack}.}
\label{tab:runtime}
\end{table}

\subsection{Evaluation for \paretosubmoddiam}
\label{sec:results-diameter}

\spara{Baselines.}
To contextualize performance for the \diameter\ cost, we include
three baselines.
Each generates a set of candidate solutions capturing different points along the
utility--diameter tradeoff; we retain only the non-dominated solutions.

\emph{\distancegreedy} iteratively selects the element maximizing marginal utility scaled by average distance to the current solution set.

\emph{\prunegraph} initializes with $\groundset$ and iteratively removes the node that causes the smallest decrease in objective value, producing a sequence of candidate tradeoffs.

\emph{\topkdegree} selects nodes in decreasing order of graph degree and returns all prefixes. 

\spara{Pareto Frontier Quality.}
Figure~\ref{fig:diameter-pareto-frontiers} shows the resulting
utility--diameter tradeoffs.
Across datasets, \cgreedydiameter\ returns high-quality
frontiers, performing slightly  better on team
formation instances and markedly better on \yelpphoenix and \yelpvegas datasets.
The \topkdegree\ heuristic is the weakest overall, while \prunegraph\ performs
competitively on \freelancer and \bibsonomy instances.

We report frontier sizes in Table~\ref{tab:frontier-size-diameter}. 
On team formation datasets, all methods yield succinct frontiers, reflecting a
limited number of diameter--utility tradeoffs;
\cgreedydiameter\ typically returns the smallest or near-smallest frontiers (e.g.,
$1.6$--$2.8$ solutions), and the baselines return comparably-sized frontiers.
In contrast, \yelpphoenix and \yelpvegas datasets exhibit extremely dense frontiers due to the combinatorial nature of pairwise distances.
We observe here that \cgreedydiameter\ is particularly effective, recovering substantially richer
frontiers (e.g., over $950$ solutions on \yelpphoenix\ and $1400$ on
\yelpvegas), consistently outperforming competing heuristics.
On the other hand, {\paretosummary} returns significantly smaller representative frontiers, with mean frontier sizes ranging from \(3.2\)–\(8.1\) points for larger datasets.
On influence maximization datasets, frontier sizes are again modest, but
\cgreedydiameter\ continues to expose the broadest set of Pareto-optimal tradeoffs.
Overall, \cgreedydiameter\ adapts to instance complexity, yielding compact
frontiers on simpler problems and rich frontiers when the diameter landscape is
dense, while {\paretosummary} consistently returns a small representative frontier even for the larger \yelpphoenix\ and \yelpvegas\ datasets 

\begin{table}
\centering
\resizebox{\linewidth}{!}{
\begin{tabular}{lccccc}
\toprule
\textbf{Dataset}
& \textbf{\cgreedydiameter}
& \prunegraph
& \distancegreedy
& \topkdegree 
& \paretosummary \\
\midrule
\freelancer & \textbf{2.8} & 2.5 & 15.1 & 2.0 & 2.1\\
\bibsonomy & \textbf{1.6} & 9.6 & 1.6 & 4.0 & 1.5\\
\imdbone & \textbf{2.1} & 2.5 & 2.0 & 2.0 & 1.9\\
\imdbtwo & \textbf{1.7} & 3.0 & 1.7 & 2.0 & 1.6\\
\yelpphoenix  & \textbf{961.1} & 15.0 & 776.7 & 9.5 & 6.8\\
\yelpvegas & \textbf{1495.2} & 19.9 & 1149.9 & 7.1 & 8.1\\
\NetPHY   & \textbf{8.5} & 2.5 & 3.0 & 3.0 & 3.2\\
\NetHEPT  & \textbf{26.5} & 3.0 & 3.0 & 4.0 & 5.3\\
\bottomrule
\end{tabular}}
\caption{Mean size of the approximate Pareto frontier produced by each algorithm for {\paretosubmoddiam}.}
\label{tab:frontier-size-diameter}
\end{table}


\spara{Running Time.}
Table~\ref{tab:runtime-diameter} reports running times for algorithms in the
{\paretosubmoddiam} setting.
Across all datasets, \cgreedydiameter\ is consistently among the fastest methods,
reflecting its reliance on a single pass over metric balls rather than repeated
optimization over constrained subproblems. {\paretosummary} also performs comparably but is up to $50\%$ slower across all datasets.
On team formation datasets, all methods run quickly, but \prunegraph\ incurs
noticeable overhead as instance sizes grow (e.g., $1.03$s on \imdbtwo\ versus
$0.014$s for \cgreedydiameter).

On \yelpphoenix and \yelpvegas, running times increase for all methods due to the
density of pairwise distances, yet \cgreedydiameter\ remains competitive
($9.83$s on \yelpphoenix, $37.5$s on \yelpvegas), matching or outperforming
\distancegreedy\ and substantially improving over \prunegraph.
The performance gap becomes more pronounced on influence maximization datasets,
where \cgreedydiameter\ scales favorably ($1.52$s on \NetPHY\ and $4.73$s on
\NetHEPT), while \prunegraph\ and \distancegreedy\ incur very large
overheads.

Overall, these results show that \cgreedydiameter\ and {\paretosummary} achieve a strong balance between frontier quality and computational efficiency, making them practical for large-scale instances where alternative heuristics do not scale.

\begin{table}
\centering
\resizebox{\linewidth}{!}{
\begin{tabular}{lccccc}
\toprule
\textbf{Dataset}
& \textbf{\cgreedydiameter}
& \prunegraph
& \distancegreedy
& \topkdegree
& \paretosummary \\
\midrule
\freelancer & \textbf{0.003} & 0.019 & 0.008 & 0.003 & 0.009\\
\bibsonomy & \textbf{0.008} & 0.659 & 0.003 & 0.042 & 0.010\\
\imdbone & \textbf{0.007} & 0.189 & 0.003 & 0.014 & 0.009\\
\imdbtwo & \textbf{0.014} & 1.031 & 0.006 & 0.066 & 0.019\\
\yelpphoenix  & \textbf{9.829} & 11.948 & 10.301 & 0.813 & 10.213\\
\yelpvegas & \textbf{37.534} & 37.877 & 30.055 & 2.904 & 32.514\\
\NetPHY & \textbf{1.521} & 152.318 & 382.095 & 1.474 & 1.928\\
\NetHEPT& \textbf{4.727} & 665.215 & 868.750 & 13.330 & 6.723\\
\bottomrule
\end{tabular}}
\caption{Running time comparison (in seconds) of algorithms for
{\paretosubmoddiam}.}
\label{tab:runtime-diameter}
\end{table}

%% file: 7conclusion.tex
\section{Conclusion}
We examined utility–cost tradeoffs in submodular maximization through the lens of Pareto optimization and introduced $(\alpha_1,\alpha_2)$-approximate Pareto frontiers. We formalized the general {\paretosubmodcost} problem and analyzed it across multiple classes of submodular objectives and cost functions arising in recommender systems, influence maximization, and team formation. 
We developed efficient, practical algorithms with provable guarantees for both the {\paretosubmodknapsack} and {\paretosubmoddiam} problems.
We demonstrated empirically that our algorithms compute high-quality approximate Pareto frontiers on real-world datasets. Together, our theoretical results and experiments establish a principled and practical framework for understanding utility–cost tradeoffs in submodular optimization. 
Our departure from the single-solution paradigm provides a conceptual shift and practical framework for understanding utility -- cost tradeoffs in submodular maximization.